\documentclass[fleqn,usenatbib]{mnras}
\pdfoutput=1

\usepackage{newtxtext,newtxmath}

\usepackage[T1]{fontenc}
\usepackage{ae,aecompl}
\usepackage{hyperref}
\usepackage{url}

\usepackage{graphicx}	
\usepackage{amsmath}	
\usepackage{amssymb}	
\usepackage{float}
\usepackage{comment}

\newcommand{\SO}{MAXI~J1727--203}

\title[MAXI J1727--203 as seen by NICER]{X-ray Spectral and Timing evolution of {\SO} with \textit{NICER}}

\author[K. Alabarta et al.]{
K. Alabarta$^{1,2}$\thanks{E-mail: k.alabarta@soton.ac.uk},
D. Altamirano$^{1}$,
M. M\'endez$^{2}$,
V. A. C\'uneo$^{3,4}$,
L. Zhang$^{1}$,
\newauthor
R. Remillard$^{5}$,
A. Castro$^{1,6}$,
R.~M.~Ludlam$^{7}$,
J. F. Steiner$^{8}$,
T. Enoto$^{9,10}$,
\newauthor
J. Homan$^{11,12}$,
Z. Arzoumanian$^{14}$,
P. Bult$^{13, 14}$,
K. C. Gendreau$^{14}$,
C. Markwardt$^{14}$,
\newauthor
T. E. Strohmayer$^{14}$,
P. Uttley$^{15}$,
F. Tombesi$^{13,14,16}$
and 
D.J.K. Buisson$^{1}$
\\
$^{1}$School of Physics and Astronomy, University of Southampton, Southampton, SO17 1BJ, UK\\
$^{2}$Kapteyn Astronomical Institute, University of Groningen, PO Box 800, NL-9700 AV Groningen, the Netherlands\\
$^{3}$Instituto de Astrof\'isica de Canarias (IAC), V\'ia L\'actea s/n, La Laguna 38205, S/C de Tenerife, Spain \\
$^{4}$Departamento de Astrof\'isica, Universidad de La Laguna, La Laguna, E-38205, S/C de Tenerife, Spain \\
$^{5}$MIT Kavli Institute for Astrophysics and Space Research, MIT, 70 Vassar Street, Cambridge, MA 02139, USA \\
$^{6}$Consorcio de Investigaci\'on del Golfo de M\'exico, CICESE, Carretera Ensenada-Tijuana 3918, 22860 Ensenada, BC, Mexico\\
$^{7}$Cahill Center for Astronomy and Astrophysics, California Institute of Technology, Pasadena, CA 91125, USA\\
$^{8}$Center for Astrophysics | Harvard \& Smithsonian, 60 Garden St. Cambridge, MA 02138, USA \\
$^{9}$Extreme Natural Phenomena RIKEN Hakubi Research Team, RIKEN Cluster for Pioneering Research, 2-1 Hirosawa, Wako,\\ Saitama 351-0198, Japan \\
$^{10}$The Hakubi Center for Advanced Research, Kyoto University, Kyoto 606-8302, Japan \\
$^{11}$Eureka Scientific, Inc., 2452 Delmer Street, Oakland, CA 94602, USA \\
$^{12}$SRON, Netherlands Institute for Space Research, Sorbonnelaan 2, 3584 CA Utrecht, The Netherlands \\
$^{13}$Department of Astronomy, University of Maryland, College Park, MD 20742, USA \\
$^{14}$Astrophysics Science Division, NASA Goddard Space Flight Center, Greenbelt, MD 20771, USA \\
$^{15}$Anton Pannekoek Institute for Astronomy, University of Amsterdam, Science Park 904, 1098 XH Amsterdam, The Netherlands\\
$^{16}$Department of Physics, University of Rome "Tor Vergata", Via della Ricerca Scientifica 1, I-00133 Rome, Italy
}

\date{Accepted 2020 July 20. Received 2020 July 20; in original form 2020 April 8}

\pubyear{2019}
\begin{document}
\sloppy
\label{firstpage}
\pagerange{\pageref{firstpage}--\pageref{lastpage}}
\maketitle

\begin{abstract}
We present a detailed X-ray spectral and variability study of the full 2018 outburst of {\SO} using \textit{NICER} observations. The outburst lasted approximately four months. Spectral modelling in the 0.3--10 keV band shows the presence of both a soft thermal and a hard Comptonised component. The analysis of these components shows that {\SO} evolved through the soft, intermediate and hard spectral states during the outburst. We find that the soft (disc) component was detected throughout almost the entire outburst, with temperatures ranging from $\sim$0.4 keV, at the moment of maximum luminosity, to $\sim$0.1 keV near the end of the outburst. The power spectrum in the hard and intermediate states shows broadband noise up to 20 Hz, with no evidence of quasi-periodic oscillations. We also study the rms spectra of the broadband noise at $0.3-10$ keV of this source. We find that the fractional rms increases with energy in most of the outburst except during the hard state, where the fractional rms remains approximately constant with energy. We also find that, below 3 keV, the fractional rms follows the same trend generally observed at energies $>3$ keV, a behaviour known from previous studies of black holes and neutron stars. The spectral and timing evolution of {\SO}, as parametrised by the hardness-intensity, hardness-rms, and rms-intensity diagrams, suggest that the system hosts a black hole, although we could not rule out a neutron star.
\end{abstract}






\section{Introduction}

Low mass X-ray binary systems (LMXBs) are binaries that contain a compact object, either a black hole (BH) or a neutron star (NS) and an evolved low-mass companion star. LMXBs for which the compact object is a black-hole candidate are known as BH low-mass X-ray binaries (BH LMXBs). The energy spectra of BH LMXBs are characterised by two main components: a soft thermal component and a hard power-law like component \citep[e.g.][]{Remillard06,Belloni10}. The thermal component is generally described by a multi-colour disc blackbody model \citep{Mitsuda84} peaking at 1--2 keV \citep[see review by ][and references therein]{Done07} and thought to be produced by a geometrically thin and optically thick accretion disc \citep{Shakura73}. The hard component is thought to be produced by a region of hot plasma, around the compact object and the accretion disc \citep[the so-called  "corona"; ][]{Sunyaev79,Sunyaev80}. A thermal Comptonisation model, in which high-energy photons are emitted by inverse Compton scattering \citep{Sunyaev80}, has been proposed to explain the hard component of the BH LMXBs energy spectra \citep[e.g.][]{Titarchuk94,Zdziarski04,Done07,Burke17}.

BH LMXBs show a variety of spectral and timing properties during an outburst \citep[see e.g.][]{VanDerKlis89,Mendez97,VanDerKlis00,Homan05,Remillard06,Belloni10,Belloni11,Plant14,Motta16}. Two main spectral states can be defined \citep[see e.g., ][]{Tanaka89, vanderKlis94}: the low/hard state (LHS), when the thermal Comptonised component dominates the energy spectrum, and the high/soft state (HSS) when the thermal component dominates the spectrum. In the LHS, however, a multi-colour disc blackbody component can be detected \citep[e.g.][]{Capitanio09a, Wang18}. In this state, the power-density spectrum (PDS) is characterised by a strong broadband noise component with a fractional rms amplitude of 30\%--50\% \citep[e.g.][]{Mendez97,Belloni05,Remillard06, Munoz-Darias11, Motta16}. In addition, quasi-periodic oscillations (QPOs) of type-C can be detected \citep[e.g., ][]{Casella04, Belloni05}. These oscillations have a centroid frequency ranging from 0.01 Hz to 30 Hz. In the HSS, a weak power-law component is sometimes detected in the energy spectrum \citep[e.g.][]{Capitanio09a}. The broadband fractional rms of BHs in this state is generally less than 5\% \citep{Mendez97}, and QPOs are sometimes detected, too \citep[e.g.][]{Remillard06,Munoz-Darias11,Motta16}.

Between the LHS and HSS, two intermediate states can be distinguished in terms of variability: the hard intermediate state (HIMS) and the soft intermediate state (SIMS) \citep[see, e.g.][]{Homan05,Belloni10}. The HIMS shows less broadband fractional rms than the hard state \citep[10\%--30\%; e.g.][]{Munoz-Darias11,Motta12} and type-C QPOs can be present \citep[e.g.][]{Casella04,Belloni05,Belloni14}. The SIMS is characterised by a weak power-law noise component that replaces the broadband noise component present in the HIMS, and type-A or type-B QPOs \citep[e.g.][]{Wijnands99,Homan01,Casella04,Belloni05,Belloni14}. Type-B QPOs have centroid frequencies in the 1--7 Hz frequency range \citep{Gao17} and a quality factor, Q > 6. Type-A QPOs have centroid frequencies in the 6.5--8 Hz frequency range and are broader than type-B and type-C QPOs, with a quality factor of Q$=$1--3 \citep{Wijnands99,Casella04,Belloni14}.

The evolution of a BH LMXB through an outburst can be well illustrated using the hardness-intensity diagram \citep[HID; e.g.][]{Homan01,Remillard06,Belloni06}. At the beginning of the outburst, the source is in the LHS and its intensity increases at approximately constant hardness ratio, drawing a vertical line in the right part of the HID. At some point, in the outburst evolution, the source starts a transition to the HSS, moving to the left in the diagram at an approximately constant luminosity. This transition corresponds to the top horizontal branch in the HID (HIMS and SIMS), reaching the HSS at the top left part of the HID. During the HSS, the source starts to decrease its intensity, moving down in the diagram. Eventually, the source returns to the HIMS and SIMS, drawing a horizontal branch in the HID, but in the opposite direction, from left to right. Before the end of the outburst, the source reaches the hard state again, to finally return to quiescence. This very particular pattern in the HID of BH LMXBs is known as the q-track and it is often discussed in terms of hysteresis \citep[e.g., ][]{Miyamoto95}. Multiple outbursts of different sources follow this q-track: e.g. XTE~1550--564, GX~339--4, H1743--322 and GRO~J1655--40   \citep{Homan01,Belloni05,Fender09,Dunn10,Uttley15}.

The outburst evolution can also be analysed using the hardness-rms diagram \citep[HRD, ][]{Belloni05} and the rms-intensity diagram \citep[RID, ][]{Munoz-Darias11}. The different spectral states show different fractional rms-hardness ratio correlations. Observations in the LHS are located on the top right side of the HRD. When the source enters the HIMS and the SIMS, it moves to the bottom left side of the HRD diagonally until the source reaches the HSS. Finally, the evolution reverses, returning to the HIMS and the SIMS following the same track as before, until it reaches the hard state again at the top right side of the HRD. The evolution in the RID is counterclockwise, similar to the one observed in the HID. In the LHS, the source evolves along a diagonal line from the bottom left to the top right of the diagram. This line is called the ``Hard Line'' \citep[HL, ][]{Munoz-Darias11}. When it makes the transition to the HIMS and SIMS, the source moves horizontally to the left side of the RID. Then the source reaches the HSS and starts to move down along a diagonal following the 1\% rms line. Finally, the source returns to the HIMS and SIMS moving horizontally to the right side of the diagram. At some point the source reaches again the 30\% rms line and goes down diagonally following the so-called ``Adjacent Hard Line'' (AHL), which is coincident to the Hard Line.

LMXBs in which the compact object is a NS are known as NS LMXBs. The energy spectra of NS LMXBs are characterised by three components: a disc blackbody component and a Comptonised component as for BH LMXBs, and a blackbody component from the emission of the surface of the NS and its boundary layer \citep[e.g.][]{Mitsuda84,DiSalvo00,Gierlinski02,Lin07}. NS LMXBs show different X-ray spectral states \citep[for a review, see ][]{vanderklis06}. At high accretion rates, NS LMXBs follow Z-tracks in the HID and the colour-colour diagrams. These sources are known as Z sources. At low accretion rates, NS LMXBs are known as atoll sources due to the tracks they follow in colour-colour diagrams \citep{Hasinger89}. Atoll sources show three X-ray spectral states that are comparable to the X-ray spectral states of BH LMXBs \citep[e.g.][]{vanderklis06, Darias14}. Besides, the hysteresis observed in BH LMXBs has been also observed in NS LMXBs \citep{Darias14}, sometimes even the q-track \citep[][]{Kording08}.

Some differences between BH LMXBs and NS LMXBs have been observed in the X-ray spectral and timing properties. The hard state of NS systems is softer than that of BH systems \citep[e.g.][]{Done03}. In terms of timing properties, the most important difference between the two types of LMXBs is the presence of kilo-hertz QPOs (kHz QPOs) at frequencies between 300 Hz and 1.2 kHz for NS \citep[][]{vanderklis06, vanDoesburgh18}. In terms of the broadband noise component and low-frequency QPOs (LFQPOs), NSs and BHs systems can be very similar \citep[e.g.][]{Klein-Wolt08} but, while BH LMXBs usually show broadband noise up to 500 Hz, NS systems can show broadband noise at higher frequencies \citep{Sunyaev00}.

%

{\SO} was discovered on 2018 June 5 with \textit{MAXI}/GSC \citep{Yoneyama18}. \citet{Ludlam18} and \citet{Kennea18} reported, respectively, observations performed the same day with the \textit{Neutron star Interior Composition Explorer} \citep[\textit{NICER}; ][]{Gendreau12} and with the Neil Gehrels \textit{Swift} Observatory \citep[\textit{Swift};][]{Gehrels04}. A hard-to-soft state transition and the disc properties of the system in the soft state, led to the possible identification of the source as a BH transient \citep{Negoro18}. In mid-July of 2018, a soft-to-hard transition was observed with \textit{Swift}/XRT \citep{Tomsick18}.


\textit{NICER} \citep{Gendreau12} is an X-ray instrument aboard the International Space Station (ISS) launched in 2017. It consists on 52 functioning detectors. Photons in the $0.2-12$ keV energy band can be detected to a time resolution of 300 ns. In this paper, we present the first study of the spectral and timing evolution of the 4-months long outburst of {\SO} as observed with \textit{NICER}.~In Section 2 we describe the observations and data analysis. In section 3 we present the results of the spectral and timing study. In section 3.1 we describe the outburst evolution. In sections 3.2 and 3.3 we describe the timing and spectral properties, respectively.  Finally, in Section 4 we discuss the nature of the compact object of the source and the identification of its spectral states. 

\section{Observation and data analysis}


\textit{NICER} observed {\SO} 86 times between 2018 June 5 and 2018 October 7 (ObsID $1200220101-1200220186$).  The data were analysed using the software HEASOFT version 6.26 and NICERDAS version 6.0. The latest CALDB version 20190516 was used.
We applied standard filtering and cleaning criteria, including the data where the pointing offset was $<54\arcsec$, the dark Earth limb angle was $>15\degr$, the bright Earth limb angle was $>30\degr$, and the International Space Station (ISS) was outside the South Atlantic Anomaly (SAA). 
We removed data from detectors 14 and 34 which occasionally show episodes of increased electronic noise, so all our results are based on using \textit{NICER}'s 50 other active detectors. 
Also, we excluded time intervals showing strong background flare-ups, that is, time intervals with an averaged count rate in the 13--15 keV energy band higher than 1 c/s. The good time intervals (GTIs) of each observation were separated into several data segments ($1-9$) based on the orbit of the ISS. The background was calculated using the ``3C50\_RGv5" model provided by the {\it \textit{NICER}} team.

To create the long-term light curve and the HID of the outburst, we first produced 1-s binned light curves in the $0.5-12$ keV, $2-3.5$ keV and $6-12$ keV energy bands for each data segments using XSELECT. We then applied the background correction for each light curve and calculated averages per data segment. We defined intensity as the average count rate in the $0.5-12$ keV energy range and the hardness ratio as the ratio between the $6-12$ keV and the $2-3.5$ keV band count rates (both background subtracted).




We extracted a background-subtracted energy spectra for each data segment using the ``3C50\_RGv5" model mentioned above. We fitted the energy spectra of {\SO} in the energy band $0.3-10$ keV using XSPEC \citep[V. 12.10.1;][]{Arnaud96}. We rebinned the spectra by a factor of 3 to correct for energy oversampling and then to have at least 25 counts per bin. In addition, we added a systematic error of 1\% in the energy range $2-10$ keV (suggested by the \textit{NICER} team). We found strong instrumental residuals below 2 keV. These residuals are typical for X-ray missions and Si-based detectors \citep[e.g.][]{Ludlam18a,Miller18}. We therefore added a 5\% systematic error in the $0.3-2$ keV energy band (also suggested by the \textit{NICER} team). We fitted the energy spectra with an absorbed \citep[\textsc{tbabs} in XSPEC, ][]{Wilms00} power-law model, \textsc{tbabs$\times$powerlaw}, an absorbed disc blackbody \citep{Mitsuda84}, \textsc{tbabs$\times$diskbb}, and an absorbed combination of a thermally Comptonisation model \citep{Zdziarski96, Zycki99} and a multi-colour disc blackbody \textsc{tbabs$\times$(nthcomp+diskbb)}. Fitting the spectra with the models \textsc{tbabs$\times$powerlaw} and \textsc{tbabs$\times$diskbb} did not give satisfactory fits in terms of $\chi^2/dof$ and expected spectral parameters. Therefore, in this paper, we only report the results of using the model \textsc{tbabs$\times$(nthcomp+diskbb)}. In order to obtain the fluxes of the different components, we added two \textsc{cflux} components to the models. The solar abundances were set according to \citet{Wilms00} and the hydrogen column density ($N_{\rm H}$) of the \textsc{tbabs} was left free. The cross section was set according to \citet{Verner96}. The 1$\sigma$ errors of the parameters were calculated from a Markov Chain Monte Carlo of length 10000 with a 2000-step burn-in phase.

For the Fourier timing analysis, we constructed Leahy-normalized power spectra \citep{Leahy83} using data segments of 131 seconds and a time resolution of 125$\mu$s. The minimum frequency was 0.007 Hz and the Nyquist frequency was 4096 Hz. Then we averaged the power spectra per observation and subtracted the Poisson noise based on the average power in the $3-4$ kHz frequency range. Finally, we converted the power spectra to squared fractional rms \citep{vanderKlis95a}. We obtained the integrated fractional rms amplitude from 0.01 Hz to 64 Hz. To obtain the rms spectrum (i.e. fractional rms amplitude vs energy), we repeated the previous procedure for the following energy bands: $0.3-0.8$ keV, $0.8-2.0$ keV, $2.0-5.0$ keV and $5.0-12$ keV. We obtained the $0.01-64$ Hz fractional rms amplitude for all these bands and plotted the fractional rms amplitude vs energy to study the evolution of the energy dependence of the fractional rms amplitude. 

To fit the power spectra we used a multi-Lorentzian function: the sum of several Lorentzians. We give the frequency of the Lorentzians in terms of the characteristic frequency, which is the frequency where the component contributes most of its variance per logarithmic interval of frequency \citep{Belloni02}: $\nu_{max}=\sqrt{\nu_{0} + (FWHM/2)^2} = \nu_{0} \sqrt{1+1/4Q^{2}}$. The quality factor $Q$ is defined as $Q = \nu_{0}/FWHM$, where $FWHM$ is the full width at half maximum and $\nu_{0}$ the centroid frequency of the Lorentzian.

\section{Results}

\subsection{Outburst evolution}

\begin{figure*}
    \centering
    \includegraphics[width=\textwidth]{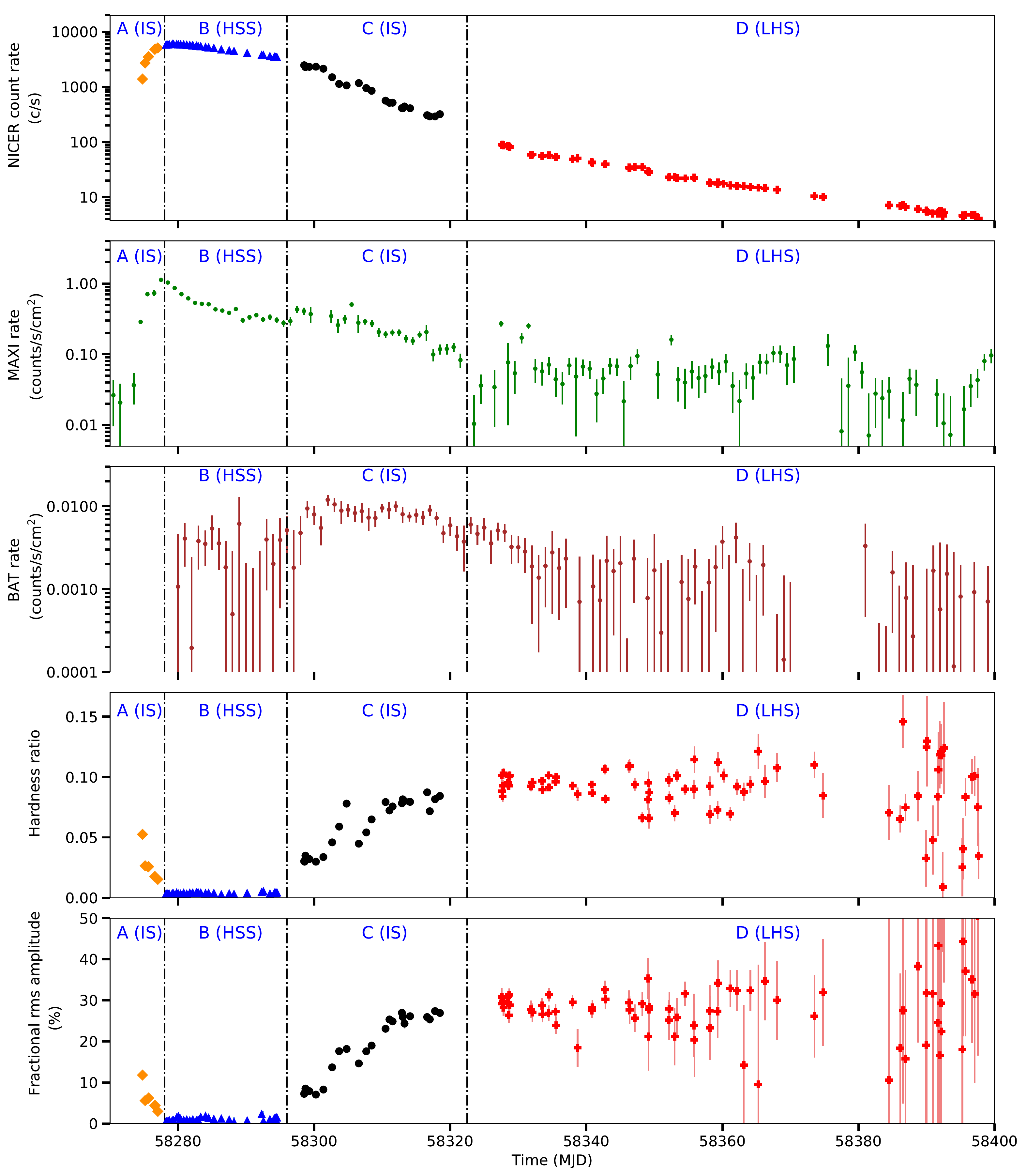}
    \caption{Top panel: \textit{NICER} light curve of the 2018 outburst of {\SO} in the $0.5-12$ keV energy band. Second panel: \textit{MAXI} light curve of the 2018 outburst of {\SO} in the $2-10$ keV energy band. Third panel: \textit{Swift}/BAT light curve in the $15-50$ keV energy band. Fourth panel: Temporal evolution of the hardness ratio ($6-12$ keV)/($2-3.5$ keV). Bottom panel: Temporal evolution of the $0.01-64$ Hz fractional rms amplitude in the $0.5-12$ keV energy band. Colours and symbols in the \textit{NICER} light curve, hardness ratio and fractional rms amplitude represent different phases of the outburst. Orange diamonds: Phase A. Blue triangles: Phase B. Black circles: Phase C. Red filled crosses: Phase D. The dotted dashed lines divide the four phases (see section 4 for a physical interpretation of these intervals).}
    \label{fig:lc}
\end{figure*}

\begin{figure}
    \centering
    \includegraphics[width=\columnwidth]{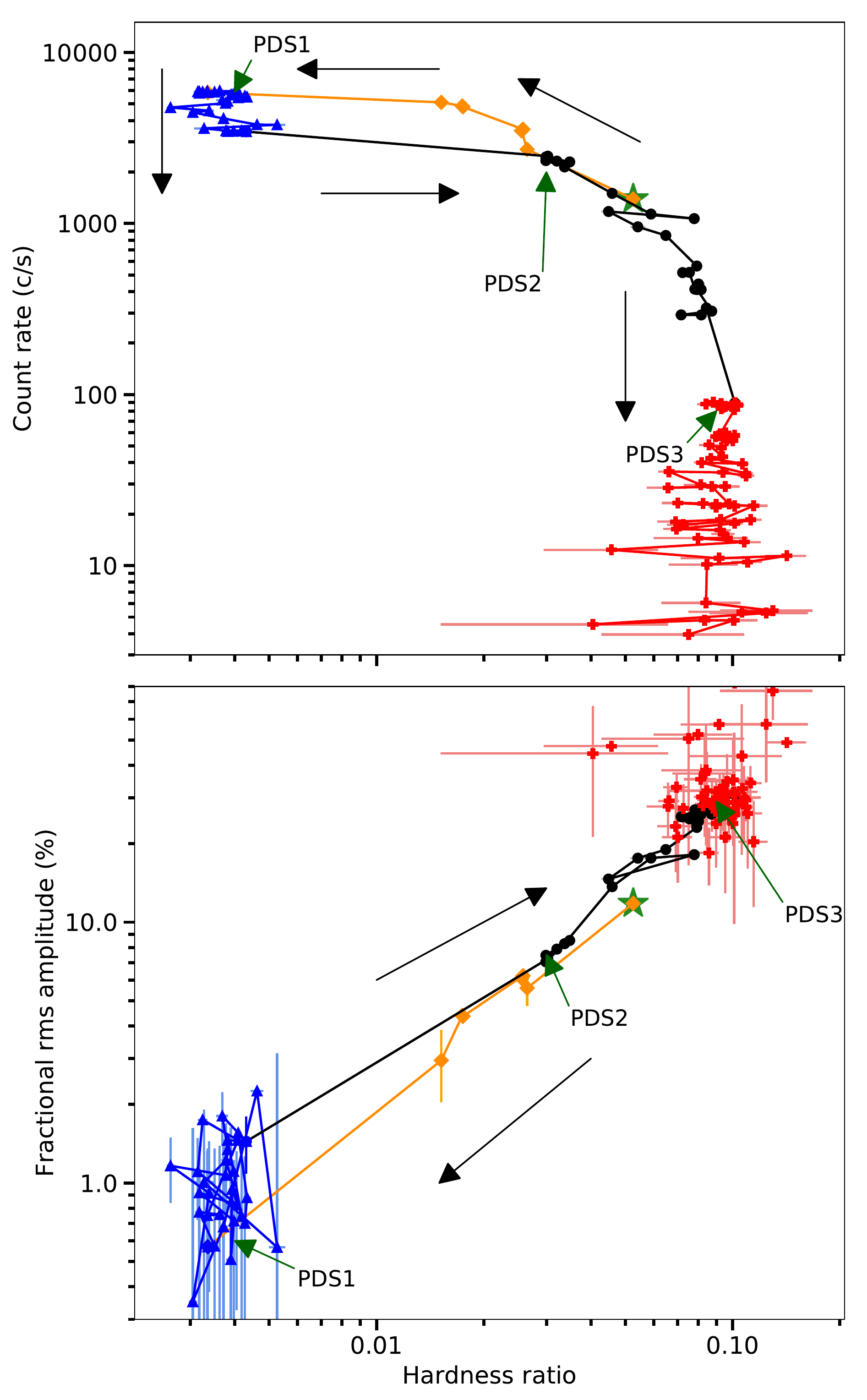}
    \caption{Top panel: HID of {\SO} during the 2018 outburst. The hardness ratio is defined as the count-ratio in the bands ($6-12$ keV)/($2.0-3.5$ keV). The count rate is obtained in the $0.5-12$ keV energy band. Bottom panel: HRD of {\SO} during the 2018 outburst. The fractional rms amplitude is obtained in the 0.01--64 Hz frequency range and $0.5-12$ keV energy band. The colours are the same as in Figure \ref{fig:lc}. The green star marks the first observation. Black arrows indicate the direction of the temporal evolution through the diagram. ``PDS1'', ``PDS2'' and ``PDS3'' mark the location in the plots of the data used to create the representative power spectra shown in Figure \ref{fig:power_example}. Colours and symbols are the same as described in Figure \ref{fig:lc}.}
    \label{fig:hid}
\end{figure}

We show the \textit{NICER} light curve of the 2018 outburst of {\SO}, which covers a period of $\sim$123 days, in the $0.5-12$ keV energy band in the top panel of Figure \ref{fig:lc}. 
Based on our spectral and variability studies, which are described below, we define four phases of the outburst in Figure \ref{fig:lc}: phase A (from MJD 58274 to MJD 58278; shown with orange diamonds), phase B (from MJD 58278 to MJD 58296; shown with blue triangles), phase C (from MJD 58296 to MJD 58322; shown with black circles) and phase D (from MJD 58322 to the end of the outburst; shown with red filled crosses). 
%

The first \textit{NICER} detection of the source was on MJD 58274, at a count rate of $\sim$1400 c/s, indicating that the \textit{NICER} observations caught the outburst already at a high flux. The intensity increased very quickly until MJD 58279 in phase A when the source reached a maximum intensity of $\sim$5960 c/s. 
After the maximum, the flux decreased monotonically, although the decay can be divided into 3 parts.
As the source entered phase B of the outburst, its intensity decreased smoothly from $\sim$5960 c/s to $\sim$3400 c/s over the next 18 days of observation.
During phase C of the outburst, the intensity decreased faster than in phase B, from $\sim$2500 c/s to $\sim$300 c/s over 26 days. Finally, in phase D the intensity decreased from $\sim$90 c/s to $\sim$4 c/s over the last 75 days of X-ray monitoring. 
After that, the apparent position of the source was located behind the Sun from the point of view of \textit{NICER} in its Earth orbit on-board the ISS. After the {\SO} occultation by the Sun, \textit{NICER} did not perform further observations of this source.

We also show the $2-10$ keV \textit{MAXI}\footnote{\href{http://maxi.riken.jp/pubdata/v6m/J1728-203/index.html}{http://maxi.riken.jp/pubdata/v6m/J1728-203/index.html}} and the $15-50$ keV \textit{Swift}/BAT\footnote{\href{https://swift.gsfc.nasa.gov/results/transients/weak/MAXIJ1727-203.lc.txt}{https://swift.gsfc.nasa.gov/results/transients/weak/MAXIJ1727-203.lc.txt}} light curves of {\SO} in the second and third panels of Figure \ref{fig:lc}, respectively.
The rise of the outburst was detected by \textit{MAXI}, showing that the intensity increased by a factor of $\sim$50 in four days. 
In phase B, the MAXI intensity decayed faster than the \textit{NICER} intensity. The analysis of the $2-10$ keV \textit{NICER} light curve  shows the same trend as the $0.5-12$ keV light curve,  probably indicating that the difference between \textit{NICER} and \textit{MAXI}  is due to differences in their respective effective areas.
In phases C and D of the outburst, the evolution of both \textit{NICER} and \textit{MAXI} light curves were similar. 
The \textit{Swift}/BAT light curve did not sample the rise of the outburst, however, it gives additional information during phases B and C, where the $15-50$ keV intensity showed a bump. In phases C and D, the \textit{Swift}/BAT intensity decayed until the end of the outburst. 

In the fourth panel in Figure \ref{fig:lc} we show the temporal evolution of the hardness ratio (as estimated from \textit{NICER} data) during the whole outburst. The different phases of the outburst show a different behaviour of the hardness ratio. In phase A of the outburst, the hardness ratio drop from $\sim$0.05 to $\sim$0.02. In phase B the hardness ratio remained constant with values around $\sim$0.004. In phase C the hardness ratio increased from $\sim$0.03 to $\sim$0.09. In phase D the source showed an approximately constant hardness ratio with an average value of $\sim$0.09. 
Due to the data-gaps between phases B and C, and phases C and D, we arbitrarily chose the limits between phases in the middle of the gap. 

In the top panel of Figure~\ref{fig:hid} we show the HID. The first point is marked with a green star in the phase A of the outburst (orange diamonds). During this phase, the source evolved in the top part of the HID from the right side to the left side. In the phase B of the outburst (blue triangles) the source reduced its intensity at an approximately constant hardness ratio. Then the source entered the phase C of the outburst (black circles) and evolved from the top left to the right side of the diagram. Finally, during the phase D of the outburst (red crosses) the source evolved to the bottom right side of the diagram. Although we are missing the rise of the outburst, a q-track shape is clear in Figure \ref{fig:hid}.

\subsection{Timing properties}

As expected from LMXBs in outburst, the X-ray variability of {\SO} also evolved through the 2018 outburst.
The bottom panel in Figure \ref{fig:lc} shows the temporal evolution of the averaged fractional rms amplitude. During the first observation (MJD 58274) {\SO} showed a fractional rms amplitude of $\sim$12\%. Then, the fractional rms decreased down to $\sim$3\% in phase A of the outburst. In phase B, the fractional rms amplitude ranged from $\sim$2\% to $\sim$0.5\%, in phase C it increased from $\sim$7\% to $\sim$27\% and, in phase D, it remained approximately constant at $\sim$30\%.

In the bottom panel of Figure~\ref{fig:hid} we show the HRD. The first observation is marked with a green star and it showed a hardness ratio of $\sim$0.05 and a fractional rms amplitude of $\sim$12\%. Then, the source evolved to the bottom left part of the diagram reaching values of the fractional rms amplitude $<1$\%. Finally, the evolution reversed and the source moved to the top right side of the diagram, increasing its hardness ratio and the fractional rms amplitude.

\begin{figure}
    \centering
    \includegraphics[width=\columnwidth]{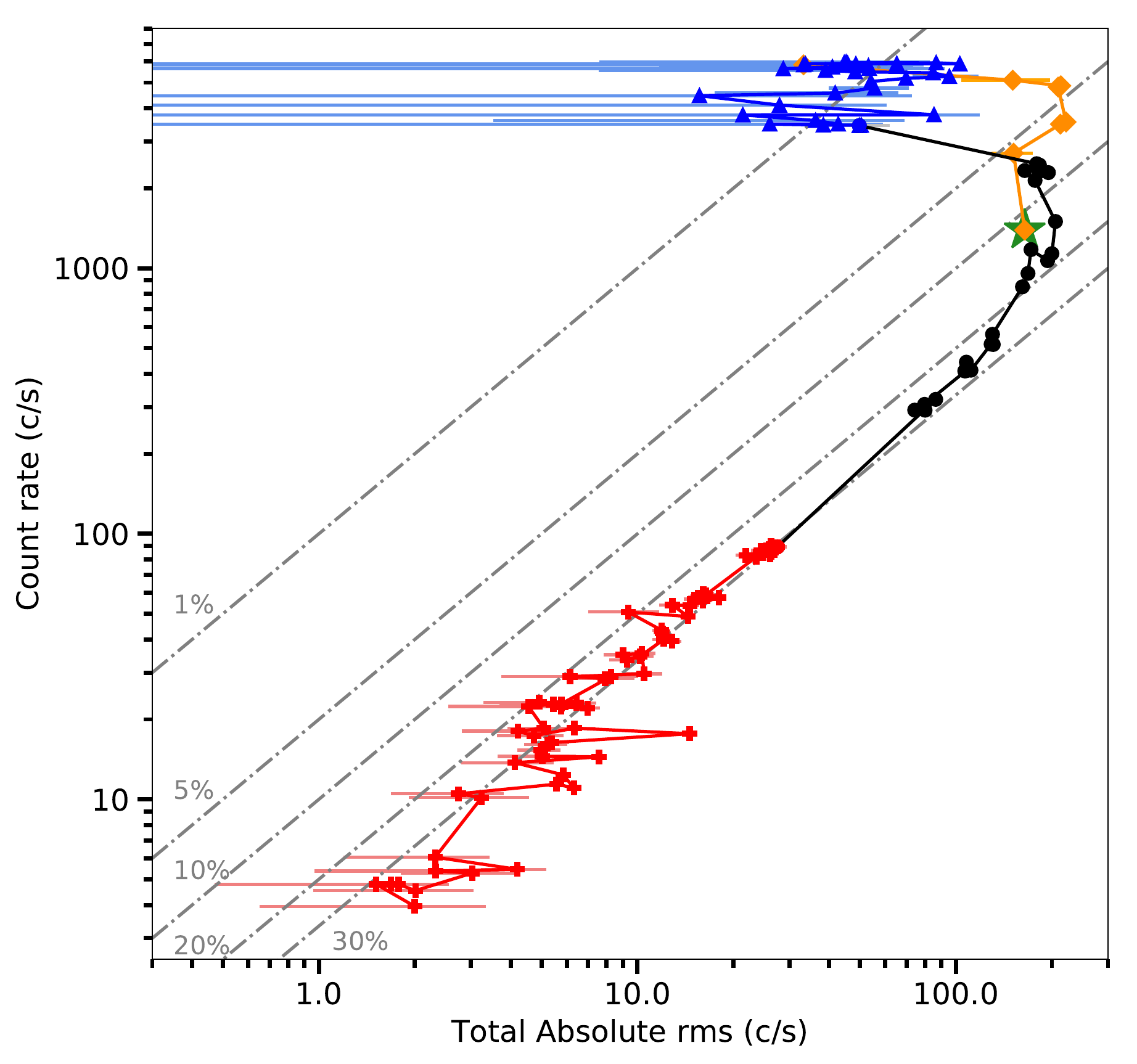}
    \caption{RID of {\SO} during the 2018 outburst. The total absolute rms is obtained in the $0.01-64$ Hz frequency range and $0.5-12$ keV energy band. The count rate is obtained in the $0.5-12$ keV energy band. The dashed lines represent the lines of constant fractional rms amplitude. Colours and symbols are the same as in Figure \ref{fig:lc}. The green star marks the first \textit{NICER} observation.}
    \label{fig:rid}
\end{figure}

Figure \ref{fig:rid} shows how {\SO} evolves in the absolute rms-intensity diagram. The source described the anticlockwise pattern that has been observed for other BHs \citep[e.g. MAXI~J1348--630, Zhang et al. in prep; GX~339--4, ][]{Munoz-Darias11}. The first point of the outburst in the RID is denoted with a green star. As the source evolved it crossed the 10\% fractional rms amplitude line increasing its intensity.  Two days later, on MJD 58276, the source crossed the 5\% line and after that it moved horizontally to the left of the diagram. From MJD 58278 to MJD 58294, {\SO} stayed close to the 1\% rms line while the intensity decreased. This corresponds to the softest part of the HID and the bottom-left part of the HRD (shown with blue triangles in the bottom panel of Figure \ref{fig:lc} and Figure \ref{fig:hid}). After MJD 58298 the source evolved in the opposite way going back to the right side of the RID. On MJD 58302, {\SO} crossed the 10\% rms line and on MJD 58310, it crossed the 20\% rms line. 17 days later, on MJD 58327, the source evolved around the 30\% rms line, identifying this as the Adjacent Hard Line. This is shown with red filled crosses in Figure \ref{fig:lc} and Figure \ref{fig:hid}.

\begin{figure*}
\centering
\includegraphics[width=\textwidth]{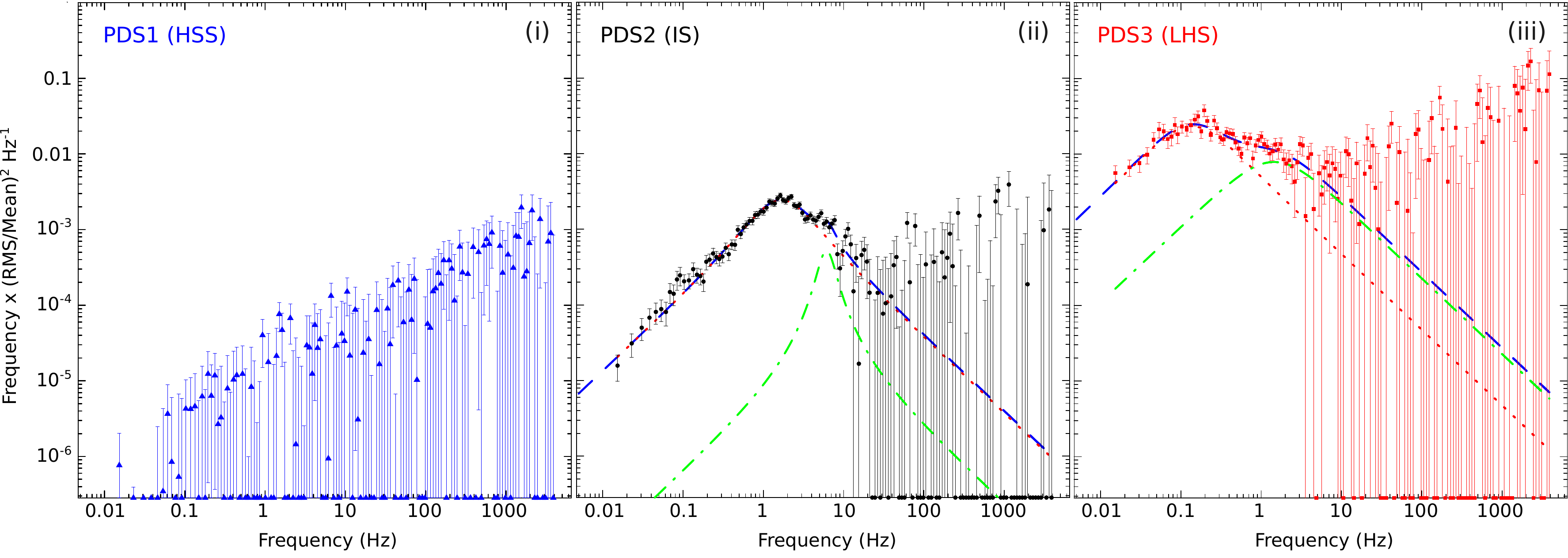}
\caption{Three representative power spectra of the 2018 outburst of {\SO}. Panel (i) shows the power spectra of ObsID 1200220105. Panel (ii) shows the power spectra of ObsID 1200220120. Panel (iii) shows the power spectra of ObsID 1200220141. These observations occurred during phases B and C, respectively, of the outburst evolution. Dashed and dotted lines represent the best fit Lorentzians.}
\label{fig:power_example}
\end{figure*}

\begin{figure}
\centering
\includegraphics[width=\columnwidth]{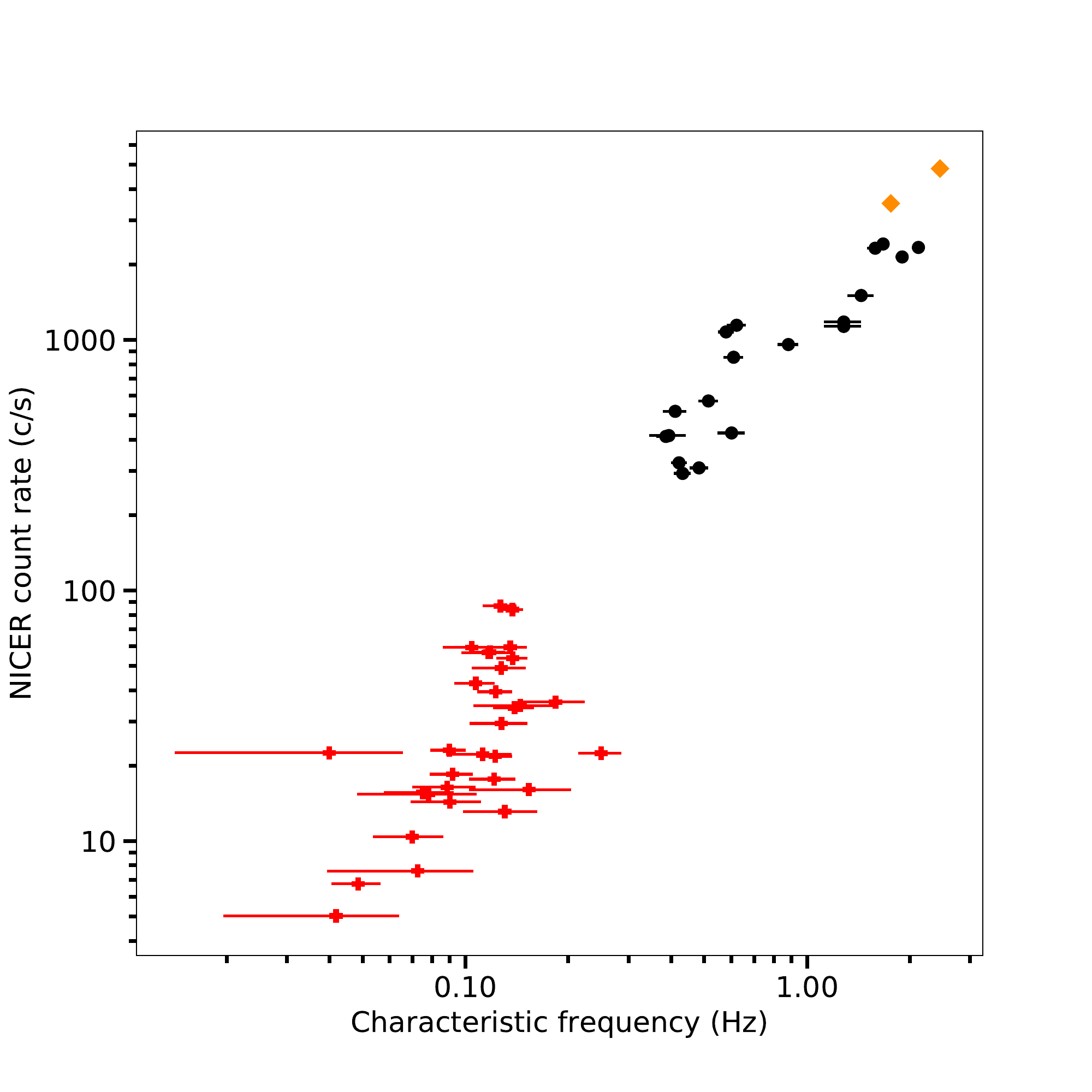}
\caption{Characteristic frequency vs \textit{NICER} count rate in the $0.5-12$ keV energy band. Colours and symbols correspond to the phases of the outburst as described in Figure \ref{fig:lc}.}
\label{f:fbreak}
\end{figure}

Figure \ref{fig:power_example} shows three representative examples of the \textit{NICER} PDS at three different stages of the outburst (marked in the HID and the RID as ``PDS1'', ``PDS2'' and ``PDS3'', respectively). The PDS of the observations before MJD 58278 (phase A of the outburst) showed a significant broadband noise component up to $\sim$20 Hz and no significant QPOs (e.g. panel (ii) in Figure \ref{fig:power_example}). In phase B, the PDS of all the observations revealed little to no significant variability (e.g. panel (i) in Figure \ref{fig:power_example}). This corresponds to the interval plotted with blue triangles of the bottom panel of Figure \ref{fig:lc} and the HID and RID. Then, from MJD 58298 to the end of the outburst (phases C and D), a broadband noise component was  present with similar power-spectral shape as that in panel (ii) and panel (iii) in Figure \ref{fig:power_example}. In this period, there was significant broadband noise extending up to a frequency of $\sim20$ Hz on MJD 58298; after this date the maximum frequency of this broadband noise component decreased down to hundredths of Hz as the source evolved towards the end of the outburst. Figure \ref{f:fbreak} shows the evolution of the characteristic frequency of the broadband noise component with intensity. We found that they are correlated. 

\begin{figure*}
\centering
\includegraphics[width=\textwidth]{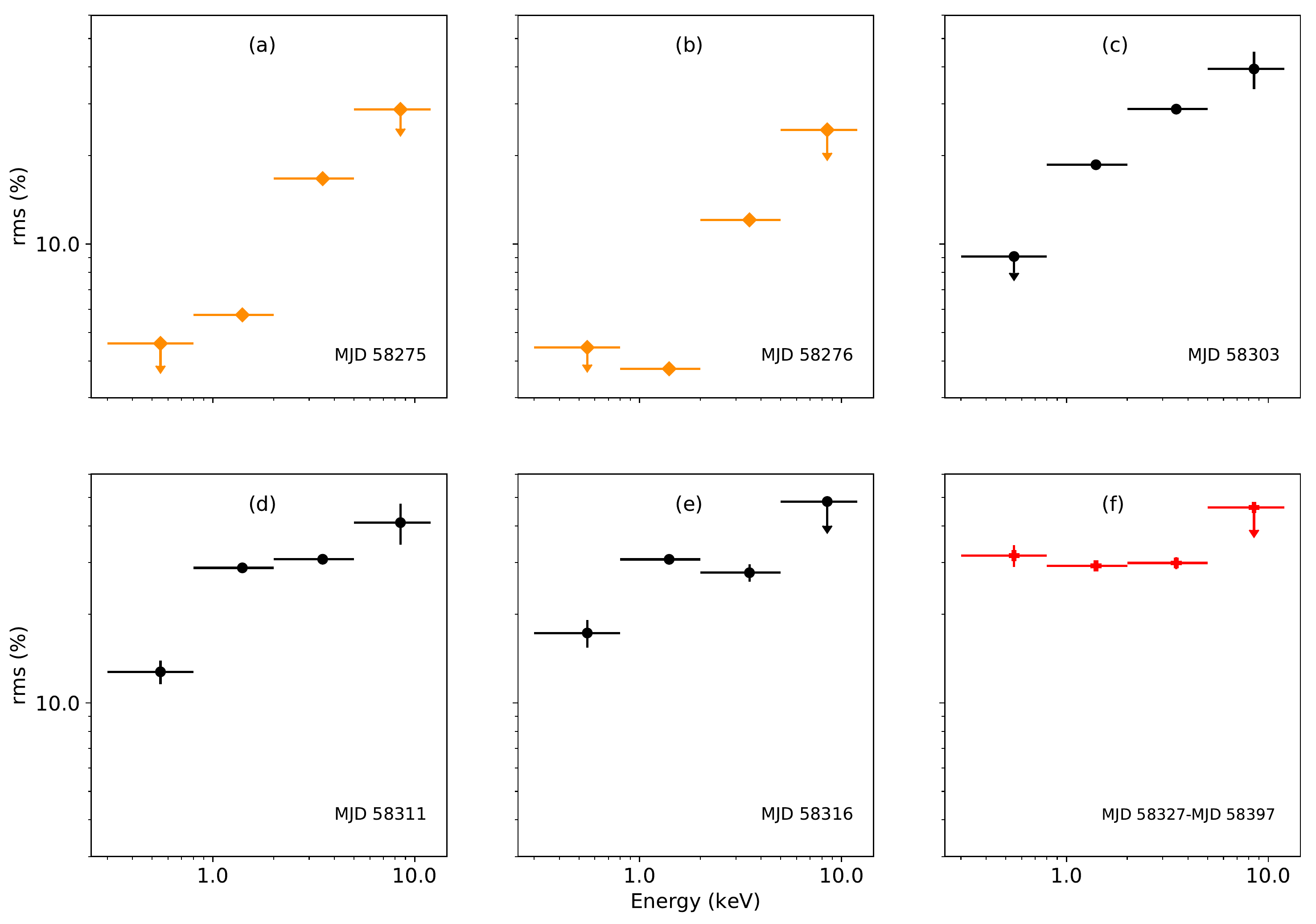}
\caption{Evolution of the $0.01-64$ Hz fractional rms spectrum of {\SO}. Plots are chronologically ordered. Colours and symbols correspond to the phases of the outburst as described in Figure \ref{fig:lc}. Arrows represent upper limits with a confidence of 3$\sigma$.}
\label{f:rms_spec}
\end{figure*}

We searched for QPOs in the PDS of {\SO} in the $0.5-12$ keV and the $2-12$ keV energy bands per observation, per orbit and per region of the HID. We found evidence for QPOs in four cases: at 0.2 Hz ($10.0\pm1.6$\% rms, ObsID 1200220134), 0.5 Hz ($7.3\pm0.7$\% RMS, ObsID 1200220127), 3 Hz ($6.1\pm0.9$\% rms, ObsID 1200220131) and 6 Hz ($1.9\pm0.2$\% rms, ObsID 1200220102). These QPOs are all between 3 and 3.5$\sigma$ significant single trial. When considering the number of trials, these QPOs are $<1\sigma$ significant; however the fractional rms amplitude we measured serve as an indication of our sensitivity to detect QPOs. 

Figure \ref{f:rms_spec} shows the $0.01-64$ Hz fractional rms spectrum of representative observations through the whole outburst in the $0.3-12$ keV energy band. The panels are chronologically ordered. The rms spectrum of panel (f) was made combining all the observations from MJD 58327 to MJD 58346 for the same reason. From panel (a) to panel (e) the fractional rms amplitude increased with energy. On panel (f) the fractional rms amplitude remained approximately constant with energy. 
During phase B of the outburst, the X-ray variability is very low, of the order of  $\sim$1 \% fractional rms. We do not show the data of this phase in Figure \ref{f:rms_spec} as we only obtain upper limits in the different energy bands.

\subsection{Spectral properties}

We fitted the energy spectra using the model \textsc{tbabs$\times$(nthcomp+diskbb)}. First, we fitted all the energy spectra separately linking the $kT_{\rm seed}$ parameter of \textsc{nthcomp} and $kT_{in}$ of \textsc{diskbb}, and we found that the electron temperature, $kT_{e}$, in \textsc{nthcomp} was always above the maximum energy of the instrument. Therefore, we fixed $kT_{e}$ at 1000 keV. Besides, we noted that the value of $N_{\rm H}$ in all the fitted energy spectra was consistent within errors. Therefore, we decided to link this parameter among all the spectra and to repeat the fitting. We obtained an average $N_{\rm H}$ of $(0.437\pm0.001)\times 10^{22}$ cm$^{-2}$, and a relatively good fit, with a $\chi^2/dof$ of 1.16, for 11107 degrees of freedom (for a total of 60 spectra). 
The excess in $\chi^2$ is given by the fit to some spectra where the 5\% of systematic errors below 2 keV were not sufficient to mitigate the effect of instrumental residuals below 2 keV. 
%

\begin{figure}
    \centering
    \includegraphics[width=\columnwidth]{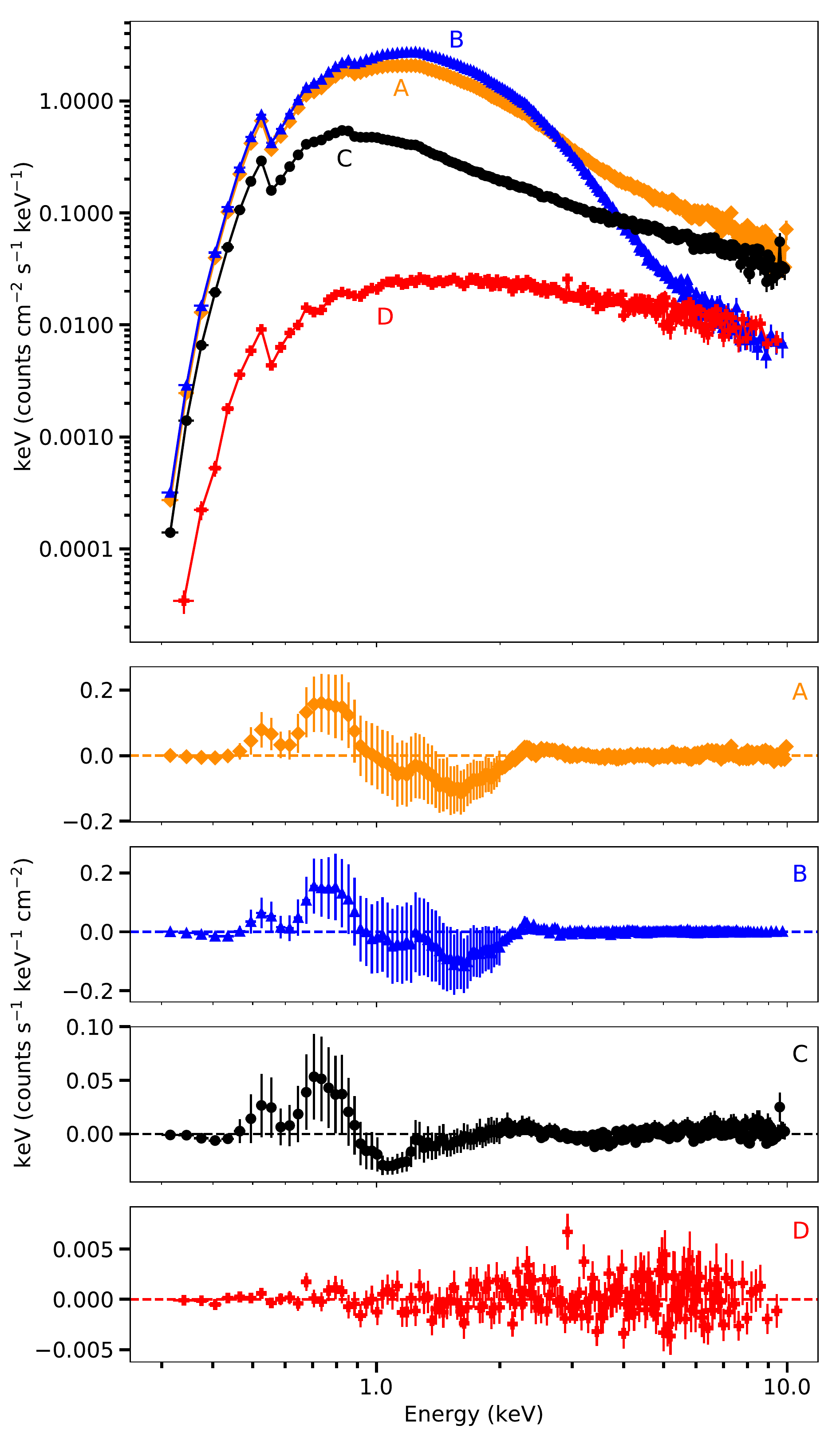}
    \caption{Upper panel: Representative energy spectra  corresponding to the different phases of the outburst. Colours and symbols represent the different phases as described in Figure \ref{fig:lc}. Dashed lines represent the best fit model in each case. Panel A, B, C and D: Residuals of the energy spectra corresponding to observations of phase A, phase B, phase C and phase D of the outburst, respectively.}
    \label{fig:spec}
\end{figure}

\begin{figure*}
    \centering
    \includegraphics[width=0.99\textwidth]{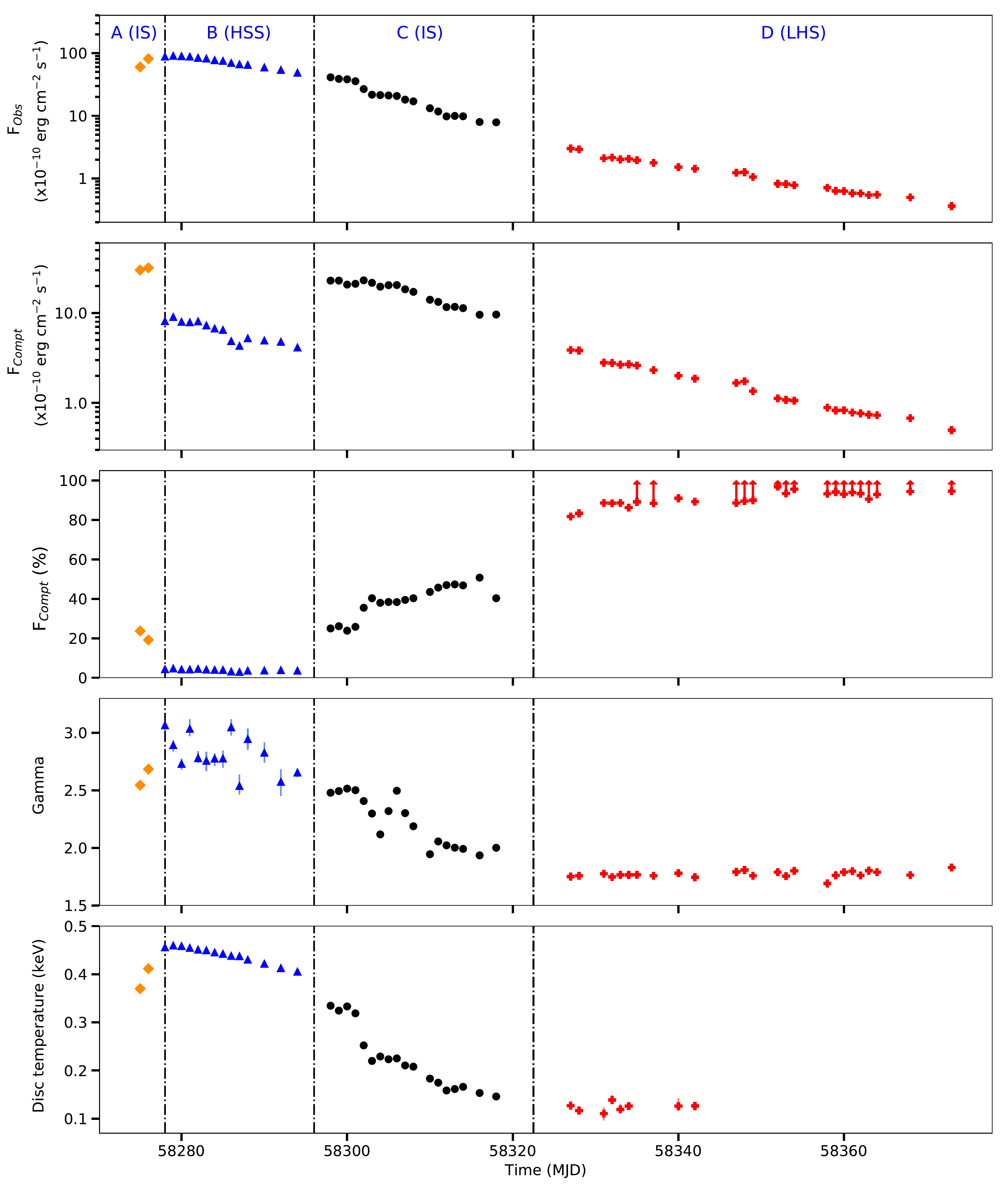}
    \caption{Evolution of the spectral parameters of {\SO} in the $0.3-10$ keV energy band. We fit \textit{NICER} X-ray spectra using a \textsc{tbabs$\times$(nthcomp+diskbb)} model. 
    From top to bottom, we plot the total observed flux, 
    the unabsorbed Comptonised flux ($F_{Compt}$), 
    the contribution of the Comptonised component to the total unabsorbed flux (\textsc{$F_{Compt}$} (\%)), 
    the photon index of \textsc{nthcomp} and the temperature at the inner disc radius of \textsc{diskbb}. After MJD 58342 the \textsc{diskbb} component is not statistically required. The values of $F_{Compt}$ (\%) after MJD 58342 are not 100\% as those take into account the contribution to the flux of a disc (at 95\% upper limits). The red arrows on the third panel represent the lower limits of the $F_{Compt}$ (\%) in observations where the disc was not significantly detected. The different colours represent the different phases of the outburst as defined in the previous Figures. The dashed line points separate the different phases.}
    \label{fig:spec_params}
\end{figure*}

Figure \ref{fig:spec} shows four representative spectra of each phase of the outburst.
The best-fitting parameters are given in table \ref{Tab:spec_params}, the evolution of the parameters is shown in Figure \ref{fig:spec_params}. 
In the upper panel of Figure \ref{fig:spec_params} we show the temporal evolution of the total observed flux. Naturally, we observed the same trend as in the upper panel of Figure \ref{fig:lc}. In the second panel of Figure \ref{fig:spec_params} we show the temporal evolution of the Comptonised component unabsorbed flux in the $0.3-10$ keV energy band. The third panel shows the contribution of the Comptonised component to the unabsorbed flux in per cent. Finally, in the last two panels, we show the temporal evolution of the photon index, $\Gamma$, of the Comptonised component and the inner disc temperature, $kT_{in}$, of the disc component. 
 
The phases identified in Figure \ref{fig:lc} show different spectral behaviour as well, as it is shown in Figure \ref{fig:spec_params}. In phase A, the contribution of the Comptonised component was $\sim$20\%. The photon index ranged from $\sim$2.5 to $\sim$2.7 and the disc temperature was close to $\sim$0.4 keV. The flux of the Comptonised component was $\sim$30.0$\times10^{-10}$ erg cm$^{-2}$s$^{-1}$. The flux of the disc component, on the other hand, ranged from $\sim$95$\times10^{-10}$ erg cm$^{-2}$ s$^{-1}$ to $\sim$135$\times10^{-10}$ erg cm$^{-2}$ s$^{-1}$. This phase corresponds to the regions with orange diamonds in Figures \ref{fig:hid} and \ref{fig:rid}. In phase B the Comptonised flux dropped to $\sim$8$\times10^{-10}$ erg cm$^{-2}$ s$^{-1}$ and decreased until $\sim$4$\times10^{-10}$ erg cm$^{-2}$ s$^{-1}$ on MJD 58294 and, as a consequence, the contribution of the Comptonised component decreased to $\sim$4\%. The photon index varied from $\sim$2.7 to $\sim$3.1 and the disc temperature decreased from $\sim$0.45 to $\sim$0.3 keV. This phase corresponds to the region plotted with blue triangles in the HID and the interval with lower fractional rms amplitude in the bottom panel of Figure \ref{fig:lc} and Figure \ref{fig:rid} (also plotted with blue triangles on the RID). In phase C the contribution of the Comptonised component was higher than in the previous phase. At the beginning of the phase, the contribution of the Comptonised component was $\sim$25\% and increased up to $\sim$40\%. The photon index and the disc temperature decreased from $\sim$2.5 to $\sim$2.0 and from $\sim$0.3 keV to $\sim$0.15 keV, respectively. This region corresponds to the black circles in the HID and the RID, where the hardness ratio and the fractional rms amplitude increased again. Finally, in phase D, the contribution of the Comptonised component to the total unabsorbed flux was higher than 80\%, with the disc component becoming insignificant (i.e. not statistically required) after MJD 58342. The photon index and the disc temperature remained approximately constant around $\sim$1.8 and $\sim$0.1 keV, respectively. This phase corresponds to the red filled crosses of the HID and the RID. In these phases, the hardness ratio and the fractional rms amplitude remained constant at their highest values.


Our spectral modelling did not require the addition of a line-component in the in the $6-7$ keV region. The addition of a Gaussian in this energy range led to non-physical results (the Gaussian component became too broad and the \textsc{nthcomp} component changed). If the sigma parameter of the line was fixed to the arbitrary value of 0.3, we found that in some cases there was a significant line. In phases A and C we could find emission lines at $\sim$6.5 keV at a significance of no more than $4\sigma$ and an equivalent width of $\sim$0.09 keV. After averaging all data of phase D in the period MJD 58327-58340, we were able to find a $\sim3\sigma$  emission line at $\sim$6.5 keV and an equivalent width of $\sim$0.05 keV (in this case the sigma parameter was also fixed to 0.3). These results suggest the potential presence of an emission line; however our results are not conclusive.

\begin{figure*}
    \centering
    \includegraphics[width=\textwidth]{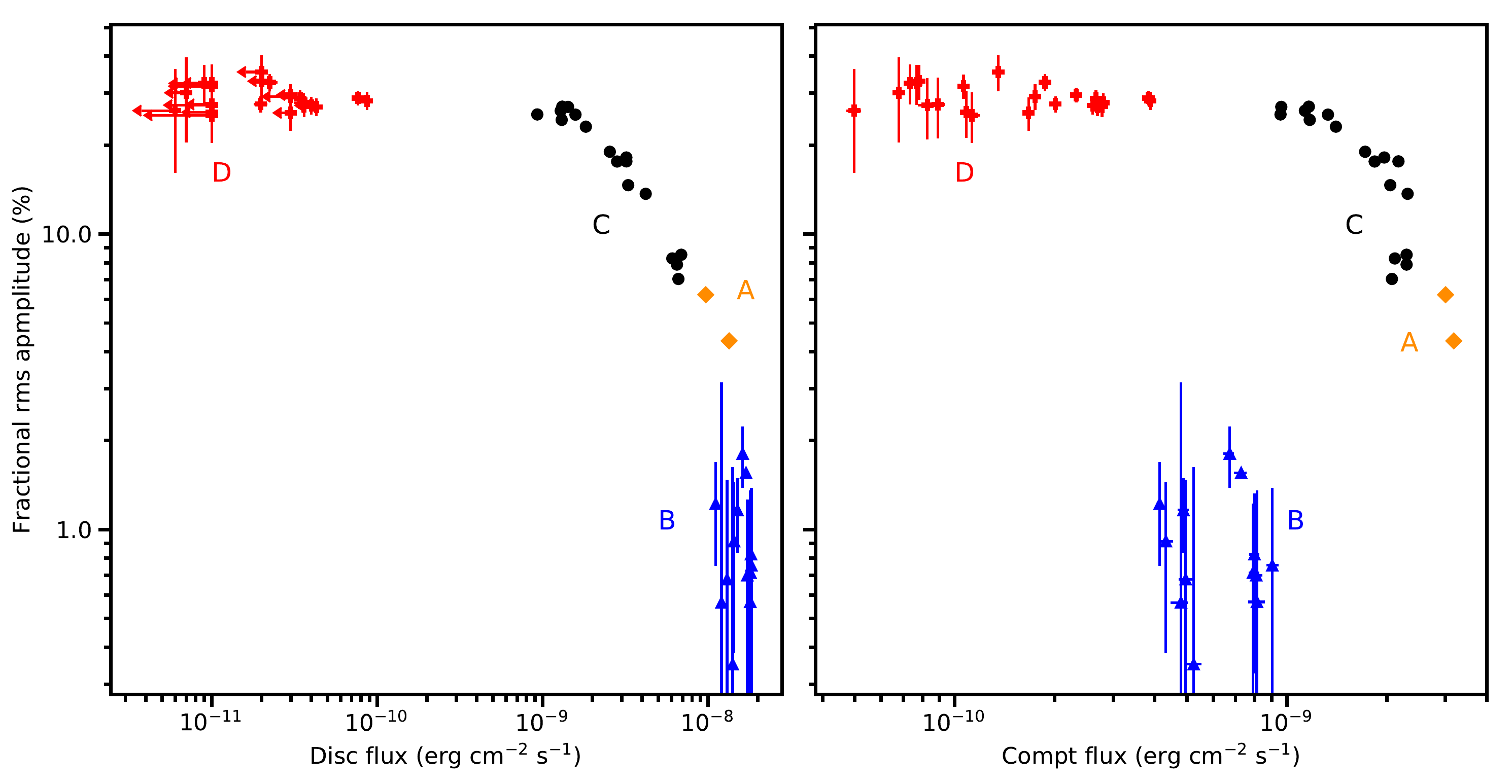}
    \caption{Plot of the $0.01-64$ Hz fractional rms amplitude ($0.5-12$ keV) versus the flux of the disc component for {\SO}. Colours and symbols represents the different phases of the outburst as in the previous plots. Red arrows represent the 95\% upper limits of the disc flux.}
    \label{fig:rms-gamma}
\end{figure*}

In Figure \ref{fig:rms-gamma} we plot the fractional rms amplitude vs. the flux of the disc component (left panel) and the fractional rms amplitude vs. the flux of the Comptonised component (right panel). 
In the left panel of Figure \ref{fig:rms-gamma}, as the disc flux increases, the fractional rms amplitude initially remains constant, phase D, and then decreases as the disc flux increases further, C, A and B.
While during phase B the rms amplitude is consistent with being constant, those measurements are consistent with the overall trend of the rms amplitude with disc flux, and extend the anti-correlation shown by the measurements in phases C and A. On the contrary, when we plot the rms amplitude vs. the Comptonised flux (right panel of Figure \ref{fig:rms-gamma}), the relation is not continuous since in those cases phase B is in between phase D and C (see Figure \ref{fig:spec_params}).


\section{Discussion}

In this paper we present a detailed spectral and timing study of the 2018 outburst of {\SO}. We found that the system showed three different spectral states during this outburst. Fitting the energy spectra of the source with a combination of a soft thermal component and a hard Comptonised component, we found that the photon index ranges between $\sim$1.75 and $\sim$3.1 and the temperature at the inner disc radius ranged between 0.1 keV and 0.45 keV. From MJD 58342 until the end of the outburst, the disc component is not detected. The power spectra of {\SO} showed broadband noise up to $\sim$20 Hz, without any significant QPOs. The $0.01-64$ Hz averaged fractional rms amplitude ($0.5-12$ keV) ranged from <1\% to $\sim$30\%. In addition, we found that the fractional rms amplitude increased with energy during most of the outburst, except at the end of the outburst when it remained approximately constant with energy. All these properties allow us to discuss the nature of the compact object of {\SO} and to determine the spectral states that characterise the source during the outburst. 

Before comparing our results with previous works, it is important to note that \textit{NICER} observations are sensitive in the $0.5-12$ keV range, whereas most of our understanding of LMXBs in the last two decades comes from observations done with the Principal Counter Array (PCA) in Rossi X-ray Timing Explorer \citep[\textit{RXTE};][]{Bradt93}, which was more sensitive in the $3-25$ keV range. 
\textit{NICER} observations, therefore, will be more affected by the interstellar absorption (which affects mainly the spectrum $<3$ keV) than those of \textit{RXTE}, affecting not only the energy spectra but also the colours/hardness estimated from them. 
Especially important is as well the role of the disc component of the spectra on the amplitude of the variability we detect \citep{Uttley11}. This is especially important for QPOs, but can also affect the broadband noise. So, for example, the integrated rms amplitudes we report in the previous section are likely underestimated as compared to those we would have measured in the usual \textit{RXTE}~$3-25$ keV energy band-pass. 
%
For this work, the difference in energy range probably had an impact on the q-shape loops in the HIDs and RID, as well as the correlations seen in the HRD. In the comparisons below, we at first neglect the energy range difference as we compare our results with those based on \textit{RXTE} data but then compare with  recent results based on \textit{NICER} observations. Our conclusions are not affected by the difference in the energy range used.   

\subsection{Nature of the compact object in \SO}

The nature of the compact object in {\SO} is still under debate. \citet{Negoro18} suggested that the source is a BH LMXB. However, the dynamical mass of the system has not been estimated yet, and the lack of very clear NS signatures (i.e. X-ray pulsations and thermonuclear X-ray bursts) does not allow to determine with absolute certainty the nature of the compact object. Below we use the evolution of the spectral and timing properties to investigate whether the compact object is a BH or a NS.

The track traced by {\SO} in the HID appears to trace part of the q-track, although we missed the rising part of the outburst. This hysteresis loop is typical of BH LMXBs \citep[e.g.][ and references therein]{Homan05, Remillard06, Fender09}. However, hysteresis loops have also been observed in NS LMXBs \citep[][]{Kording08, Darias14}. 

An evolution in the HRD similar to that of {\SO} has been observed in other LMXBs. The track we found for {\SO} is similar to the one showed by GX~339--4 \citep{Belloni05} and MAXI~J1348--630 (Zhang et al. in prep.), both BH LMXBs. Nevertheless, similar tracks were observed in two NS LMXBs \citep[Aql~X--1 and 4U~1705--44, ][]{Darias14}.

The track that {\SO} traced in the RID is similar to the track traced by the data of the BH candidate GX~339--4 \citep[based on \textit{RXTE}; ][]{Munoz-Darias11} and  MAXI~J1348--630 (based on \textit{NICER} data; Zhang et al. in prep). The difference between {\SO} and GX~339--4 is that the Adjacent Hard Line of the latter was located between 30\% and 40\% fractional rms amplitude, whereas in the case of {\SO} the Adjacent Hard Line line was located between 20-30\% fractional rms. 
%
%
\citet{Darias14} found that NS LMXBs also show hysteresis in the RID. In particular, these authors found that low accretion rate NS LMXBs traced similar tracks in the RID as those traced by BH LMXBs. The main difference between these low accretion rate NS systems and BH systems is that the track followed during the state transitions is diagonal in NS, while state transitions in BH are usually horizontal, at least for low-inclination systems \citep{Munoz-Darias13}. This makes low accretion rate NS brighter during the soft than during the hard or intermediate states in the $3-15$ keV energy band, as opposed to low-inclination BH LMXBs \citep{Munoz-Darias13}. In Figures \ref{fig:hid} and \ref{fig:rid}, it is observed that, unlike the transitions found by \citet{Darias14}, the transition from the right to the left part of the diagrams are horizontal, suggesting a BH nature for {\SO}.     

The X-ray timing properties of {\SO} do not allow us to determine the nature of the compact object in the system due to the lack of specific BH and NS signatures. 
The presence of kHz QPOs or X-ray pulsations and type-A, B and C QPOs would have allowed us to identify the compact object as a NS or a BH, respectively. Unfortunately, no kHz QPOs or X-ray pulsations are observed in the PDS of \SO. We found some marginally significant QPOs in the PDS at low frequencies (from 0.2 to 6 Hz); however the data are not sufficient to identify them with the NS or BH QPO counterparts.
We can focus on the maximum frequency of the variability of the broadband noise component. {\SO} showed broadband noise component that extends up to 20 Hz. Based on the results of \citet{Sunyaev00}, this behaviour is more typical of BHs, since the power spectra of BH LMXBs displays a strong decline at frequencies higher than $10-50$ Hz, with no significant variability above $100-200$ Hz (e.g. GX~339--4, GS~1354--644, XTE~J1748--288, and 4U~1630--47). NS LMXBs, on the other hand, can show significant variability in the power spectra up to 500-1000 Hz (e.g. 4U1608--522, SAX~J1808.4--3658, and 4U0614+091). The fact that {\SO} showed variability only up to 20 Hz suggests that the compact object in {\SO} is a BH. 

The evolution of the energy dependence of the fractional rms amplitude of the broadband noise component at energies $0.3-12$ keV is consistent with what has been seen in other BH LMXBs in energies above $2-3$ keV \citep[e.g. XTE~J1550--564 and XTE~J1650--50;][]{Gierlinski05}. We found that during most of the outburst, the fractional rms amplitude increased with energy. The only exception is shown in the panel (f) of Figure \ref{f:rms_spec} where the rms remained approximately constant with energy. This panel corresponds to the spectrally hardest observations in Figures \ref{fig:hid} and \ref{fig:rid} (red crosses of the diagrams). \citet{Gierlinski05} found that in the hard state of XTE~J1550--564 and XTE~J1650--50 the rms-spectra remained constant or slightly decreased with energy, while in the intermediate and the soft state the rms spectra increased with energy. Some NS show a similar behaviour \citep[e.g. XTE J1701-462;][]{Bu15}. The rms spectra of broadband noise components of NS increase with energy in some cases \citep[][studying XTE J1701--462 and MAXI J0911--655, respectively]{Bu15, Bult17}, as {\SO} did during most of the outburst. Unfortunately, neither of those works present the evolution of the rms spectra during a whole outburst, so at the moment we cannot compare the evolution of these sources and that of {\SO}.

The evolution of the spectral parameters of {\SO} is similar to what has been observed in other BH LMXBs and NS LMXBs. The photon index of {\SO} ranged from $\sim$1.75 to $\sim$3.1. This evolution is similar to two of the most studied BH LMXBs, Cyg~X--1 \citep[photon index from $\sim$1.5 to $\sim$2.7, ][]{Titarchuk94}  and GX~339--4 \citep[photon index from $\sim$1.5 to $\sim$2.9, ][]{Plant14}. We also compared the photon index of {\SO} with the photon index of two NS LMXBs: 4U 1636-53 and MXB~1658--298 during its 2015-2017 outburst. In the case of 4U 1636-53, the photon index ranged from 1.3 to 2.5 \citep{Zhang17}. This range is similar to the photon index range of \SO. In the case of MXB~1658--298, the photon index ranged from $\sim$1.7 to $\sim$2.4 \citep{Sharma18}. 
Although the photon index in NS LMXBs might show lower values than for BH LMXBs (something that would have to be tested studying a much larger sample), this potential difference would argue that {\SO} is a BH candidate. In terms of the inner disc temperature, {\SO} showed a lower temperature than other BH LMXBs. The disc temperature of GX~339--4 ranged from $\sim$0.6 to $\sim$0.9 keV \citep{Plant14} and the temperature of Cyg~X--1 ranged between 0.5 keV and 0.6 keV \citep{Shaposhnikov06}. The temperature of 4U 1636-53 ranged from $\sim$0.3 to $\sim$0.8 keV in the best-fitting results \citep{Zhang17} and the disc temperature of MXB~1658--298 ranged from $\sim$0.6 to $\sim$0.9 keV \citep{Sharma18}. The disc temperature of {\SO} was lower than these 4 systems.

A potential explanation for a lower temperature in {\SO} than in other sources could be related to the mass of the compact object. Assuming that the accretion disc is at the innermost stable circular orbit (ISCO), the temperature at the inner disc radius is proportional to $\left( \dot{M} / M^2 \right)^{1/4}$, where $\dot{M}$ is the mass accretion rate and $M$ is the mass of the compact object \citep{Frank02}. According to \citet{Darias08}, $M<6{\rm M}_{\odot}$ for GX~339--4. We take the temperature of GX~339--4 in the soft state (0.79 keV) from \citet{Plant14}, and we also take the disc temperature of the softest observation in the HID of {\SO} (0.45 keV). From that, if we assume that $\dot{M}$ is the same for two sources in the same spectral state, we estimate a lower limit for the mass of {\SO} of $\sim$19 ${\rm M}_{\odot}$.  Therefore, the high mass of the compact object can explain the low temperature of the inner disc.
Alternatively, as suggested by \citet{Gou11}, the low temperature at the inner disc radius can be a consequence of a low inclination of the accretion disc with respect to the line of sight. 

We can repeat this analysis to estimate the mass of {\SO} with the NS Aql X--1. For this, we took the temperature of Aql~X--1 in the soft state of its 2007 outburst \citep[$\sim$0.66 keV, ][]{Raichur11}. Considering a mass of $\sim1.4\;{\rm M}_{\odot}$ for the NS in Aql~X--1, we obtained a mass of $\sim2\;{\rm M}_{\odot}$ for the compact object in {\SO}. If we consider a higher mass for Aql~X--1, the mass of \SO~also increases.  Considering the latter, this mass estimates suggest that the compact object in {\SO} is massive NS or a low-mass BH.

Based on all the above comparisons, although we cannot unambiguously identify the nature of the compact object in {\SO}, the evolution in the HID, RID and RHD, and the temperature at the inner radius of the accretion disc during the softest observations,  suggest that it is a BH.


%

\subsection{Anticorrelation between the fractional rms amplitude and the flux of the disc component}

Figure \ref{fig:rms-gamma} shows that the relation between the fractional rms amplitude and the disc flux is continuous and that both quantities are anti-correlated during phases A and C (orange diamonds and black circles) of the outburst. While in phases B and D (blue triangles and red crosses) the rms amplitude is consistent with being independent of the disc flux, those measurements extend the relation seen in phases A and C to low (phase D) and high (phase B) values of the disc flux.   

The simplest interpretation of this behaviour is that the variability is produced by the Comptonised component, the disc emission is not variable and, as the relative contribution of the disc to the total emission increases, the variability decreases. If this is the case, the intrinsic variability would be produced by the corona \citep[e.g., for QPOs][]{Lee98,Lee01,Kumar14}. \citet{Karpouzas20} explain this for the kilohertz QPOs in neutron-star LMXBs, and \citet{Zhang20} for the type C QPOs in the black-hole candidate GRS 1915+105, but the same mechanism could apply for the broadband component that we discuss here, but the observed trend would be driven by the disc. A similar argument was discussed by \citet{Mendez01} for the dependence of the rms amplitude of the kilohertz QPOs in the neutron-star LMXBs 4U~1728--34, 4U~1608--52 and Aql~X--1.

\subsection{Spectral states of \SO}

Assuming that the source is a BH LMXB, we can identify its spectral states from its spectral and timing properties. Here we describe the different spectral states found for {\SO}:

\begin{itemize}
    \item \textbf{Low/hard state (LHS)}: From MJD 58327 to MJD 58397. This period corresponds with phase D of the outburst, marked with red filled crosses in Figures \ref{fig:lc}, \ref{fig:hid}, \ref{fig:rid}, \ref{fig:spec_params} and \ref{fig:rms-gamma}. In the HID the source was in the right vertical branch, with hardness values close to $\sim$0.1. The fractional rms amplitude in the LHS state was close to $\sim$30\% and followed the AHL in the RID, supporting the LHS classification state based on the results of \citet{Munoz-Darias11}. The fractional rms amplitude was also approximately constant with energy (panel with red filled crosses in Figure \ref{f:rms_spec}). The PDS was dominated by a broadband noise component (panel (iii) in Figure \ref{fig:power_example}). In terms of spectral properties, the contribution of the Comptonised component was >80\% in this state. At the end of the outburst, the disc component is not significant. The fractional rms amplitude is not correlated with the flux of the Comptonised component. This is because the fractional rms remains approximately constant with energy. The photon index of the Comptonised component and the inner disc temperature remained approximately constant at $\sim$1.8 and $\sim$0.1 keV, respectively. 
    
    \item \textbf{Intermediate states (IS)}: From MJD 58274 to MJD 58278 and from MJD 58298 to MJD 58327. These periods corresponds to phases A and C of the outburst, respectively, which are marked with orange diamonds (phase A) and black circles (phase C) in Figures \ref{fig:lc}, \ref{fig:hid}, \ref{fig:rid}, \ref{fig:spec_params} and \ref{fig:rms-gamma}. In the HID these correspond to the horizontal branches with hardness values from $\sim$0.02 to $\sim$0.1. In these periods, the fractional rms amplitude ranged from $\sim$5\% to $\sim$30\% and the source evolved to the top left part of the RID in the first epoch (MJD 58274--MJD 58278) and to the Adjacent Hard Line in the second epoch (MJD 58298--MJD 58327), as can be seen in Figure \ref{fig:rid}. The fractional rms amplitude increased with energy (panels with black circles and orange diamonds in Figure \ref{f:rms_spec}), and the PDS was dominated by a broadband noise component (panel (ii) in Figure \ref{fig:power_example}). No significant QPOs are detected during this phase. The characteristic frequency also increases with the intensity. The contribution of the Comptonised component ranged from $\sim$20\% to 50\%. The fractional rms amplitude and the flux of the Comptonised flux are anti-correlated, suggesting that the change of variability is driven by changes in the flux of the Comptonised component. The photon index of the Comptonised component in this state ranged from $\sim$2.0 to $\sim$2.7 and the inner disc temperature decreased from $\sim$0.4 keV to $\sim$0.15 keV. 
    
    \item \textbf{High/soft state (HSS)}: From MJD 58278 to MJD 58298. This period corresponds to phase B of the outburst, plotted with blue triangles in Figures \ref{fig:lc}, \ref{fig:hid}, \ref{fig:rid}, \ref{fig:spec_params} and \ref{fig:rms-gamma}. In the HID the hardness ratio was approximately constant close to $\sim$0.005. The fractional rms amplitude was $\sim$1\%, which can be observed in the top left part of the RID, where the source evolved around the 1\% fractional rms line. The maximum frequency also increases with energy. The rms spectrum increases with energy (blue triangles in Figure \ref{f:rms_spec}). The contribution of the Comptonised component was less than 5\%, the photon index of the Comptonised component ranged from $\sim$2.5 to $\sim$3.0 and the temperature of the inner disc decreased from $\sim$0.45 keV to $\sim$0.3 keV. The rms-flux correlation was flat with some scatter (Figure \ref{fig:rms-gamma}). 
\end{itemize}

\section*{Acknowledgements}

This work is based on observations made by the \textit{NICER} X-ray mission supported by NASA. This research has made use of data and software provided by the High Energy Astrophysics Science Archive Research Center (HEASARC), a service of the Astrophysics Science Division at NASA/GSFC and the High Energy Astrophysics Division of the Smithsonian Astrophysical Observatory. This research has made use of the \textit{MAXI} light curve provided by RIKEN, JAXA, and the \textit{MAXI} team. This research has also made use of \textit{Swift}/BAT transient monitor results provided by the \textit{Swift}/BAT team. KA acknowledges support from a UGC-UKIERI Phase 3 Thematic Partnership (UGC-UKIERI-2017-18-006; PI: P. Gandhi). KA especially acknowledges Dr. Keith Arnaud for his help with the X-ray tool XSPEC. D.A. and D.J.K.B. acknowledge support from the Royal Society. V.A.C. acknowledges support from the Royal Society International Exchanges "The first step for High-Energy Astrophysics relations between Argentina and UK" and from the Spanish \textit{Ministerio de Ciencia e Innovaci\'on} under grant AYA2017-83216-P. LZ and AC acknowledges support from the Royal Society Newton International Fellowship. R.M.L. acknowledges the support of NASA through Hubble Fellowship Program grant HST-HF2-51440.001.

\section*{Data availability}

The data underlying this article are publicly available in the High Energy Astrophysics Science Archive Research Center (HEASARC) at
\begin{sloppypar}
  The data underlying this article are publicly available in the High Energy Astrophysics Science Archive Research Center (HEASARC) at \url{https://heasarc.gsfc.nasa.gov/db-perl/W3Browse/w3browse.pl}
\end{sloppypar}






\bibliographystyle{mnras}
\bibliography{nicer}




\appendix

\section{Spectral fitting parameters}

\begin{table*}
\caption{Summary of the spectral parameters of {\SO} in the 0.3--10 keV energy band. Errors represent the 1$\sigma$ level confidence interval of the parameter. For the observations in which we do not detect the disc component significantly we give the 95\% confidence upper limit of the disc normalisation and the corresponding disc flux.}
\resizebox{\textwidth}{!}{
\begin{tabular}{cccccccccc}
\hline 
MJD & $\Gamma$ & \textsc{nthcomp} norm. & $kT_{in}$ & \textsc{diskbb} norm. & Comptonised flux & Disc flux & Unabsorbed flux & Total absorbed flux & Phase \\\textit{NICER}
 & & ($\times 10^{-2}$) & (keV) & [$\times 10^{4}$ km$^2$ (10 kpc)$^{-2}$] & ($\times 10^{-10}$ erg cm$^{-2}$ s$^{-1}$) & ($\times 10^{-10}$ erg cm$^{-2}$ s$^{-1}$) & ($\times 10^{-10}$ erg cm$^{-2}$ s$^{-1}$) & ($\times 10^{-10}$ erg cm$^{-2}$ s$^{-1}$) & outburst \\
\hline
\smallskip
\smallskip
58275 & 2.53$\pm0.03$ & 50.9$\pm0.1$ & 0.370$\pm0.001$ & 2.99$\pm0.03$ & 30.0$\pm0.3$ & 96.7$^{+0.6}_{-1.0}$ & 126.8$^{+0.7}_{-1.0}$ & 60.2$\pm0.2$ & A \\
\smallskip
\smallskip
58276 & 2.68$\pm0.03$ & 50.5$\pm0.2$ & 0.412$\pm0.003$ & 2.62$\pm0.08$ & 31.8$\pm0.5$ & 134.0$\pm0.9$ & 166$\pm1$ & 81.3$^{+0.4}_{-0.3}$ & A \\ 
\smallskip
\smallskip
58278 & 3.1$\pm0.1$ & 12.7$\pm0.1$ & 0.456$\pm0.002$ & 2.28$\pm0.04$ & 8.1$\pm0.5$ & 179.8$\pm0.4$ & 187.9$\pm0.6$ & 88.00$^{+0.09}_{-0.06}$ & B \\ 
\smallskip
\smallskip
58279 & 2.89$\pm0.05$ & 13.3$\pm0.1$ & 0.459$\pm0.002$ & 2.24$\pm0.04$ & 9.0$\pm0.4$ & 183.0$\pm0.8$ & 192.0$\pm0.8$ & 90.7$\pm0.3$ & B \\ 
\smallskip
\smallskip
58280 & 2.73$\pm0.05$ & 11.1$\pm0.1$ & 0.458$\pm0.002$ & 2.26$\pm0.04$ & 8.0$\pm0.3$ & 181.7$\pm0.9$ & 189.8$\pm0.9$ & 89.3$\pm0.4$ & B \\ 
\smallskip
\smallskip
58281 & 3.04$\pm0.07$ & 12.33$\pm0.08$ & 0.455$\pm0.001$ & 2.32$\pm0.02$ & 7.9$\pm0.3$ & 181.0$\pm0.8$ & 188.4$\pm0.9$ & 88.0$\pm0.3$ & B  \\ 
\smallskip
\smallskip
58282 & 2.78$\pm0.05$ & 11.71$\pm0.09$ & 0.451$\pm0.002$ & 2.29$\pm0.04$ & 8.1$\pm0.4$ & 173.0$\pm0.8$ & 181.0$\pm0.8$ & 84.2$\pm0.3$ & B  \\ 
\smallskip
\smallskip
58283 & 2.76$\pm0.09$ & 10.5$\pm0.1$ & 0.449$\pm0.002$ & 2.28$\pm0.04$ & 7.3$\pm0.4$ & 169.6$\pm0.8$ & 176.9$\pm0.9$ & 81.9$\pm0.3$ & B  \\ 
\smallskip
\smallskip
58284 & 2.78$^{+0.04}_{-0.07}$ & 9.90$\pm0.09$ & 0.445$\pm0.001$ & 2.27$\pm0.02$ & 6.7$\pm0.3$ & 161.8$\pm0.8$ & 169.0$\pm0.9$ & 77.3$\pm0.3$ & B  \\ 
\smallskip
\smallskip
58285 & 2.78$\pm0.08$ & 9.67$\pm0.07$ & 0.442$\pm0.001$ & 2.28$\pm0.02$ & 6.5$\pm0.3$ & 158.3$\pm0.4$ & 164.8$\pm0.5$ & 75.1$\pm0.2$ & B  \\ 
\smallskip
\smallskip
58286 & 3.05$\pm0.07$ & 8.06$\pm0.07$ & 0.438$\pm0.001$ & 2.27$\pm0.02$ & 4.9$\pm0.2$ & 150.5$\pm0.7$ & 155.4$\pm0.7$ & 69.8$\pm0.2$ & B  \\ 
\smallskip
\smallskip
58287 & 2.54$\pm0.08$ & 5.90$\pm0.06$ & 0.437$\pm0.001$ & 2.17$\pm0.02$ & 4.3$\pm0.2$ & 143.5$\pm0.5$ & 147.8$\pm0.6$ & 66.3$\pm0.1$ & B  \\ 
\smallskip
\smallskip
58288 & 2.95$\pm0.09$ & 8.57$\pm0.09$ & 0.430$\pm0.001$ & 2.28$\pm0.02$ & 5.2$\pm0.3$ & 140.8$\pm0.7$ & 146.0$\pm0.7$ & 64.9$\pm0.2$ & B  \\ 
\smallskip
\smallskip
58290 & 2.83$\pm0.09$ & 8.00$\pm0.08$ & 0.421$\pm0.002$ & 2.30$\pm0.04$ & 5.0$\pm0.2$ & 130.2$\pm0.7$ & 135.1$\pm0.8$ & 59.1$\pm0.3$ & B  \\ 
\smallskip
\smallskip
58292 & 2.6$\pm0.1$ & 7.2$\pm0.1$ & 0.413$\pm0.002$ & 2.34$\pm0.05$ & 4.8$\pm0.3$ & 121$\pm1$ & 125$\pm1$ & 53.9$\pm0.3$ & B  \\ 
\smallskip
\smallskip
58294 & 2.66$\pm0.04$ & 6.55$\pm0.05$ & 0.4053$\pm0.0007$ & 2.32$\pm0.02$ & 4.14$\pm0.08$ & 111.0$\pm0.6$ & 115.2$\pm0.6$ & 48.7$\pm0.2$ & B  \\ 
\smallskip
\smallskip
58298 & 2.48$\pm0.02$ & 42.1$\pm0.1$ & 0.330$\pm0.001$ & 3.29$\pm0.04$ & 22.9$\pm0.2$ & 68.8$\pm0.6$ & 91.7$\pm0.6$ & 41.1$\pm0.1$ & C \\ 
\smallskip
\smallskip
58299 & 2.49$\pm0.02$ & 43.7$\pm0.1$ & 0.324$\pm0.002$ & 3.55$\pm0.09$ & 22.9$\pm0.3$ & 64.8$\pm0.5$ & 87.8$\pm0.6$ & 38.8$\pm0.1$ & C \\ 
\smallskip
\smallskip
58300 & 2.52$\pm0.02$ & 38.9$\pm0.1$ & 0.333$\pm0.002$ & 3.23$\pm0.08$ & 20.7$\pm0.2$ & 66.2$\pm0.9$ & 86.9$\pm0.9$ & 38.2$\pm0.2$ & C  \\
\smallskip
\smallskip
58301 & 2.50$\pm0.02$ & 41.1$\pm0.1$ & 0.319$\pm0.002$ & 3.59$\pm0.09$ & 21.1$\pm0.3$ & 60.8$\pm0.4$ & 81.9$\pm0.5$ & 35.60$^{+0.08}_{-0.03}$ & C  \\ 
\smallskip
\smallskip
58302 & 2.41$\pm0.02$ & 51.2$\pm0.02$ & 0.252$\pm0.003$ & 7.1$\pm0.3$ & 23.1$\pm0.3$ & 41.9$\pm0.6$ & 65.1$\pm0.6$ & 26.7$\pm0.1$ & C  \\  
\smallskip
\smallskip
58303 & 2.30$\pm0.01$ & 48.3$\pm0.1$ & 0.220$\pm0.003$ & 10.1$\pm0.6$ & 21.7$\pm0.2$ & 32.1$\pm0.5$ & 53.7$\pm0.5$ & 21.70$\pm0.09$ & C  \\ 
\smallskip
\smallskip
58304 & 2.12$\pm0.01$ & 37.9$\pm0.1$ & 0.229$\pm0.002$ & 8.3$\pm0.3$ & 19.6$\pm0.2$ & 32.0$\pm0.4$ & 51.7$\pm0.4$ & 21.40$\pm^{+0.05}_{-0.09}$ & C  \\ 
\smallskip
\smallskip
58305 & 2.32$\pm0.02$ & 45.7$\pm0.2$ & 0.223$\pm0.003$ & 9.6$\pm0.5$ & 20.4$\pm0.3$ & 32.7$\pm0.6$ & 53.1$\pm0.6$ & 21.1$\pm0.1$ & C  \\ 
\smallskip
\smallskip
58306 & 2.50$\pm0.02$ & 50.6$\pm0.3$ & 0.225$\pm0.004$ & 9.3$\pm0.6$ & 20.5$\pm0.3$ & 32.9$\pm0.4$ & 53.4$\pm0.6$ & 20.70$\pm0.06$ & C  \\ 
\smallskip
\smallskip
58307 & 2.30$\pm0.02$ & 41.7$\pm0.2$ & 0.210$\pm0.003$ & 10.9$\pm0.6$ & 18.4$\pm0.3$ & 28.2$\pm0.4$ & 46.6$\pm0.5$ & 18.10$\pm0.08$ & C  \\ 
\smallskip
\smallskip
58308 & 2.188$\pm0.009$ & 36.4$\pm0.1$ & 0.2079$\pm0.0009$ & 10.4$\pm0.2$ & 17.20$\pm0.08$ & 25.5$\pm0.2$ & 42.7$\pm0.3$ & 17.00$\pm0.05$ & C  \\ 
\smallskip
\smallskip
58310 & 1.944$\pm0.008$ & 25.5$\pm0.1$ & 0.183$\pm0.002$ & 13.7$\pm0.6$ & 14.04$\pm0.07$ & 18.2$\pm0.2$ & 32.3$\pm0.2$ & 13.2$\pm0.06$ & C  \\ 
\smallskip
\smallskip
58311 & 2.05$\pm0.01$ & 26.5$\pm0.1$ & 0.175$\pm0.002$ & 14.9$\pm0.7$ & 13.29$\pm0.09$ & 15.8$\pm0.3$ & 29.1$\pm0.3$ & 11.70$\pm0.03$ & C  \\ 
\smallskip
\smallskip
58312 & 2.02$\pm0.02$ & 22.8$\pm0.2$ & 0.159$\pm0.002$ & 20$\pm1$ & 11.64$\pm0.07$ & 13.1$\pm0.2$ & 24.8$\pm0.2$ & 9.79$\pm0.08$ & C  \\ 
\smallskip
\smallskip
58313 & 2.00$\pm0.02$ & 22.6$\pm0.2$ & 0.161$\pm0.002$ & 18.0$\pm0.9$ & 11.7$\pm0.1$ & 13.0$\pm0.2$ & 24.7$\pm0.2$ & 9.96$\pm0.06$ & C  \\ 
\smallskip
\smallskip
58314 & 1.99$\pm0.01$ & 21.64$\pm0.08$ & 0.166$\pm0.002$ & 15.5$\pm0.8$ & 11.33$\pm0.06$ & 12.9$\pm0.3$ & 24.2$\pm0.3$ & 9.82$\pm0.04$ & C  \\ 
\smallskip
\smallskip
58316 & 1.935$^{+0.008}_{-0.011}$ & 17.53$\pm0.08$ & 0.153$\pm0.003$ & 17$\pm1$ & 9.57$\pm0.07$ & 9.3$\pm0.3$ & 18.9$\pm0.3$ & 7.97$\pm0.04$ & C  \\ 
\smallskip
\smallskip
58318 & 2.001$\pm0.009$ & 17.50$\pm0.09$ & 0.146$\pm0.004$ & 16$\pm2$ & 9.6$\pm0.1$ & 14$\pm1$ & 24$\pm1$ & 7.87$\pm0.04$ & C  \\ 
\smallskip
\smallskip
58327 & 1.75$\pm0.03$ & 6.03$\pm0.05$ & 0.127$\pm0.008$ & 4$\pm1$ & 3.88$\pm0.05$ & 0.87$\pm0.07$ & 4.75$\pm0.08$ & 3.00$\pm0.04$ & D \\ 
\smallskip
\smallskip
58328 & 1.76$\pm0.01$ & 5.97$\pm0.04$ & 0.117$\pm0.009$ & 6$\pm2$ & 3.83$\pm0.04$ & 0.77$\pm0.04$ & 4.59$\pm0.06$ & 2.92$\pm0.03$ & D  \\ 
\smallskip
\smallskip
58331 & 1.77$\pm0.02$ & 4.42$\pm0.04$ & 0.11$\pm0.01$ & 4$\pm1$ & 2.80$\pm0.06$ & 0.36$\pm0.05$ & 3.16$\pm0.07$ & 2.10$\pm0.02$ & D  \\ 
\smallskip
\smallskip
58332 & 1.75$\pm0.01$ &  4.31$\pm0.03$ & 0.139$^{+0.009}_{-0.006}$ & 1.1$\pm0.3$ & 2.78$\pm0.03$ & 0.36$\pm0.04$ & 3.14$\pm0.05$ & 2.16$\pm0.02$ & D  \\ 
\smallskip
\smallskip
58333 & 1.77$\pm0.01$ & 4.52$\pm0.03$ & 0.12$\pm0.09$ & 2.3$^{+1.0}_{-0.7}$ & 2.66$\pm0.03$ & 0.34$\pm0.04$ & 3.00$\pm0.04$ & 2.020$^{+0.006}_{-0.010}$ & D  \\ 
\smallskip
\smallskip
58334 & 1.77$\pm0.02$ & 4.24$\pm0.03$ & 0.126$\pm0.007$ & 2.1$\pm0.5$ & 2.69$\pm0.03$ & 0.43$\pm0.03$ & 3.12$\pm0.04$ & 2.06$\pm0.02$ & D  \\
\smallskip
\smallskip
58335 & 1.77$\pm0.01$ & 4.08$\pm0.02$ & -- & <5 & 2.60$\pm0.03$ & <0.4 & 2.92$\pm0.05$ & 1.95$\pm0.02$ & D  \\ 
\smallskip
\smallskip
58337 & 1.76$\pm0.02$ & 3.63$\pm0.02$ & -- & <3 & 2.32$\pm0.05$ & <0.3 & 2.62$\pm0.05$ & 1.780$^{+0.020}_{-0.007}$ & D  \\ 
\smallskip
\smallskip
58340 & 1.779$^{+0.011}_{-0.005}$ & 3.41$\pm0.02$ & 0.13$\pm0.02$ & 0.9$\pm0.3$ & 2.01$\pm0.03$ & 0.19$\pm0.02$ & 2.20$\pm0.04$ & 1.520$\pm0.006$ & D  \\ 
\smallskip
\smallskip
58342 & 1.74$\pm0.02$ & 2.89$\pm0.02$ & 0.126$\pm0.009$ & 1.1$\pm0.3$ & 1.87$\pm0.03$ & 0.22$\pm0.03$ & 2.09$\pm0.04$ & 1.43$\pm0.01$ & D  \\ 
\smallskip
\smallskip
58347 & 1.79$\pm0.02$ & 2.79$\pm0.03$ & -- & <5 & 1.67$\pm0.04$ & <0.3 & 1.88$\pm0.06$ & 1.24$\pm0.02$ & D  \\ 
\smallskip
\smallskip
58348 & 1.808$^{+0.007}_{-0.012}$ & 2.79$\pm0.02$ & -- & <19 & 1.74$\pm0.04$ & <0.3 & 1.95$\pm0.09$ & 1.260$^{+0.010}_{-0.007}$ & D  \\
\smallskip
\smallskip
58349 & 1.76$\pm0.03$ & 2.09$\pm0.03$ & - & <0.6 & 1.35$\pm0.06$ & <0.2 & 1.50$\pm0.07$ & 1.06$\pm0.03$ & D  \\ 
\hline
\end{tabular}}
\label{Tab:spec_params}
\end{table*}

\addtocounter{table}{-1}
\begin{table*}
\caption{Continued}
\resizebox{\textwidth}{!}{
\begin{tabular}{cccccccccc}
\hline 
MJD & $\Gamma$ & \textsc{nthcomp} norm. & $kT_{in}$ & \textsc{diskbb} norm. & Compt. flux & Disc flux & Unabsorbed flux & Total absorbed flux & Phase\\
 & & ($\times 10^{-2}$) & (keV) & [$\times 10^{4}$ km$^2$ (10 kpc)$^{-2}$] & ($\times 10^{-10}$ erg cm$^{-2}$ s$^{-1}$) & ($\times 10^{-10}$ erg cm$^{-2}$ s$^{-1}$) & ($\times 10^{-10}$ erg cm$^{-2}$ s$^{-1}$) & ($\times 10^{-10}$ erg cm$^{-2}$ s$^{-1}$) & outburst \\
\hline
\smallskip
\smallskip
58352 & 1.79$\pm0.03$ & 1.79$\pm0.02$ & -- & <2 & 1.13$\pm0.05$ & <0.1 & 1.16$\pm0.07$ & 0.830$^{+0.020}_{-0.009}$ & D  \\ 
\smallskip
\smallskip
58353 & 1.75$\pm0.01$ & 1.68$\pm0.01$ & -- & <2 & 1.08$\pm0.03$ & <0.1 & 1.16$\pm0.04$ & 0.815$\pm0.008$ & D  \\ 
\smallskip
\smallskip
58354 & 1.799$\pm0.009$ & 1.71$\pm0.01$ & -- & <3 & 1.06$\pm0.04$ & <0.1 & 1.11$\pm0.05$ & 0.779$^{+0.009}_{-0.004}$ & D  \\ 
\smallskip
\smallskip
58358 & 1.69$\pm0.02$ & 1.29$\pm0.01$ & -- & <2 & 0.89$\pm0.04$ & <0.1 & 0.95$\pm0.05$ & 0.710$^{+0.009}_{-0.006}$ & D  \\ 
\smallskip
\smallskip
58359 & 1.76$\pm0.04$ & 1.30$\pm0.01$ & -- & <0.6 & 0.83$\pm0.05$ & <0.1 & 0.88$\pm0.07$ & 0.64$\pm0.01$ & D  \\ 
\smallskip
\smallskip
58360 & 1.79$\pm0.02$ & 1.33$\pm0.01$ & -- & <0.5 & 0.83$\pm0.03$ & <0.1 & 0.89$\pm0.04$ & 0.627$\pm0.007$ & D  \\ 
\smallskip
\smallskip
58361 & 1.79$\pm0.02$ & 1.26$\pm0.01$ & -- & <0.8 & 0.78$\pm0.03$ & <0.02 & 0.83$\pm0.03$ & 0.582$^{+0.006}_{-0.002}$ & D  \\ 
\smallskip
\smallskip
58362 & 1.76$\pm0.03$ & 1.24$\pm0.01$ & -- & <0.7 & 0.77$\pm0.02$ & <0.09 & 0.82$\pm0.03$ & 0.579$^{+0.010}_{-0.005}$ & D  \\ 
\smallskip
\smallskip
58363 & 1.80$\pm0.02$ & 1.19$\pm0.02$ & -- & <2 & 0.74$\pm0.02$ & <0.08 & 0.82$\pm0.02$ & 0.54$\pm0.01$ & D  \\ 
\smallskip
\smallskip
58364 & 1.79$\pm0.02$ & 1.17$\pm0.01$ & -- & <1 & 0.73$\pm0.03$ & <0.1 & 0.79$\pm0.04$ & 0.549$^{+0.010}_{-0.007}$ & D  \\ 
\smallskip
\smallskip
58368 & 1.76$\pm0.03$ & 1.04$\pm0.02$ & -- & <4 & 0.68$\pm0.03$ & <0.07 & 0.72$\pm0.04$ & 0.500$^{+0.020}_{-0.004}$ & D  \\ 
\smallskip
\smallskip
58373 & 1.83$\pm0.03$ & 0.841$\pm0.009$ & -- & <0.4 & 0.49$\pm0.03$ & <0.06 & 0.53$\pm0.04$ & 0.363$^{+0.008}_{-0.003}$ & D  \\ 
\smallskip
\smallskip
58374 & 1.86$\pm0.04$ & 0.78$\pm0.01$ & -- & <0.3 & 0.46$\pm0.02$ & <0.08 & 0.50$\pm0.03$ & 0.345$^{+0.007}_{-0.010}$ & D  \\
\hline
\end{tabular}}
\label{Tab:spec_params_cont}
\end{table*}


\label{lastpage}
\end{document}